\documentclass[useAMS,usenatbib,a4paper]{mn2e}

\usepackage{times}
\usepackage{amsmath,amssymb}
\usepackage{bm}
\usepackage{color}
\usepackage{epsfig}
\usepackage{mathrsfs}
\renewcommand{\mathbf}{\bm}
\newcommand{\ba}{\begin{eqnarray}}
\newcommand{\ea}{\end{eqnarray}}

\newcommand{\omegam}{\Omega_{{m}}}

\newcommand{\omegak}{\Omega_{{k}}}
\newcommand{\omegmo}{\Omega_{{m}}}

\newcommand{\omegko}{\Omega_{{k}}}
\newcommand{\omegmin}{\Omega_{\textnormal{in}}}
\newcommand{\omegmout}{\Omega_{\textnormal{out}}}

\newcommand{\kapr}{\kappa(r)r^2}
\newcommand{\Hperp}{H_{\perp}}
\newcommand{\Hperpo}{H_{\perp_0}}
\newcommand{\Hpar}{H_{\parallel}}
\newcommand{\aperp}{a_{\perp }}

\newcommand{\apar}{a_{\parallel}}

\newcommand{\dA}{d_{\textnormal{A}}}

\newcommand{\LCDM}{\Lambda\textnormal{CDM}}
\newcommand{\hu}{kms$^{-1}$Mpc$^{-1}$}

\newcommand\aj{{AJ}}%
\newcommand\mnras{{MNRAS}}%
\newcommand\apjl{{ApJ}}%
\newcommand\prd{{Phys.~Rev.~D}}%
\newcommand\apj{{ApJ}}%
\newcommand\apjs{{ApJS}}%
\newcommand\aap{{A\&A}}%
\title{Rendering Dark Energy Void}
\author[S. February et al.]{Sean February\thanks{sean.astrofizik@gmail.com}, Julien Larena\thanks{julien.larena@gmail.com}, Mathew Smith\thanks{mathew.smith@uct.ac.za} and Chris Clarkson\thanks{chris.clarkson@uct.ac.za}\\
Astrophysics, Cosmology and Gravitation Centre, and,  Department of Mathematics
and Applied Mathematics, University of Cape Town, \\Rondebosch 7701,
Cape Town, South Africa}
\date{\today}
\begin{document}

\maketitle

\label{firstpage}

\begin{abstract}
Dark energy observations may be explained within general relativity using an inhomogeneous Hubble-scale depression in the matter density and accompanying curvature, which evolves naturally out of an Einstein-de Sitter (EdS) model. We present a simple parameterization of a void which can reproduce concordance model distances to arbitrary accuracy, but can parameterize away from this to give a smooth density profile everywhere. We show how the Hubble constant is not just a nuisance parameter in inhomogeneous models because it affects the shape of the distance-redshift relation. Independent Hubble-rate data from age estimates can in principle serve to break the degeneracy between concordance and void models, but the data is not yet able to achieve this. Using the latest Constitution supernova dataset we show that robust limits can be placed on the size of a void which is roughly independent of its shape. However, the sharpness of the profile at the origin cannot be well constrained due to supernova being dominated by peculiar velocities in the local universe.  We illustrate our results using some recently proposed diagnostics for the Friedmann models.
\end{abstract}


\section{Introduction}
\label{sec:intro}

An odd explanation for the dark energy problem in cosmology is one where the underlying geometry of the universe is significantly inhomogeneous on Hubble scales, and not homogeneous as the standard model assumes. These models are possible because we have direct access only to data on our nullcone and so can't disentangle temporal evolution in the scale factor from radial variations. Such explanations are ungainly compared with standard cosmology because they revoke the Copernican principle, placing us at or very near the centre of the universe. Perhaps this is just because the models used~-- Lema\^itre-Tolman-Bondi (LTB) or Szekeres to date~\citep{moffat92,moffat94,MHE,sugiura,alnes,celerier1,VFW,notari,celerier2,CFL,gbh1,gbh2,szekeres,yoo,zibinvoid,bolejko,CBKH,parallax}~-- are very simplistic descriptions of inhomogeneity, and more elaborate inhomogeneous ones will be able to satisfy some version of the Copernican principle (CP) yet satisfy observational constraints on isotropy (e.g., a Swiss-Cheese model~\citep{marra1, biswas,  marra2} or something like that).\footnote{See~\citet{CB} and ~\citet{BC} for examples of spacetimes which, although unrealistic, are globally inhomogeneous yet satisfy a version of the cosmological principle.} We may instead think of these models as smoothing all observables over the sky and so compressing all inhomogeneities into one or two radial degrees of freedom (d.o.f.) centred about us~-- and so we needn't think then as ourselves `at the centre of the universe' in the standard way.\footnote{As argued by K. Bolejko and M.-N. Celerier, private communication.}  Whatever the interpretation, such models are at the toy stage, and have not been developed to any sophistication beyond understanding the background dynamics, and observational relations; in particular, perturbation theory and structure formation is more-or-less unexplored, though this is changing~\citep{tomita,zibin,pertpaper}. They should, however, be taken seriously because we don't yet have a physical explanation for dark energy in any other form. Indeed, one could argue that these models are in fact the most conservative explanation for dark energy, as no new physics needs to be introduced.

Regardless of the details, these models raise an important question for cosmology, particularly so in the light of the dark energy problem: can we test the Copernican principle? While many would argue that it has effectively been done via the success and accuracy of the standard model, until we have a physical explanation for dark energy (and the inflaton for that matter) we are open to the accusation of having only phenomenological descriptions of two key observations in cosmology: accelerating expansion and scale-invariant initial conditions. Only a handful of tests have actually been proposed: the Goodman-Caldwell-Stebbins test, which looks at the Cosmic Microwave Background (CMB) inside our past lightcone~\citep{goodman,caldwell}; and the `curvature test', which checks if today's value of the curvature parameter, $\Omega_k$, given by
\ba\label{OK}
\Omega_k=\frac{\left\{H(z)[(1+z)d_A(z)]_{,z}\right\}^2-1}{[H_0(1+z)d_A(z)]^2},
\ea
 yields the same answer regardless of the redshift of measurement~\citep{CBL,UCE}, as it must in any Friedmann-Lema\^itre-Robertson-Walker (FLRW) universe. A crucial issue for this test is that it requires two independent measurements: one for distances, from, e.g. Type Ia Supernovae (SNIa), and one for the Hubble rate, from, e.g. Baryon Acoustic Oscillations (BAO) or age estimates.
However, recognising the difficulties posed by the  requirement of independent observables that these tests require, it has recently been proposed instead that we can test the Copernican principle from SNIa alone~\citep{CFL}, thereby making the process much simpler than these other methods. If one demands that the void is suitably smooth at the centre, then, as argued in~\citet{CFL}, this leads to inevitable differences from the $\Lambda$ cold dark matter ($\LCDM$) distance modulus which can be detected with future SNIa observations such as the SuperNova Acceleration Probe (SNAP).  The essence of this argument is as follows. A generic LTB model has two radial functional d.o.f.. One of these is the bang time function, which controls any inhomogeneity of the big bang surface; the other may be chosen to be the radial curvature today. These functions can be chosen such that the observer at the centre observes distances and $H(z)$ (or number-counts) exactly as one would in a $\LCDM$ model~\citep{MHE,CBKH}. (In this scenario the curvature test would have to be performed using both radial and tangential Hubble rates~-- see below for definitions.) However, if the bang time function is not constant then this excites modes which are decaying~\citep{silk}; hence, if this is  significant enough a factor at late times to affect these observables the models would have to be outrageously inhomogeneous at early times~-- this doesn't rule them out a priori, but does mean such models would have to have their early time behaviour rigourously examined. One can argue on the basis of this that a `realistic' LTB model has only one true d.o.f.. Now, if this d.o.f. is chosen to reproduce the distance modulus of a $\LCDM$ model, the void profile must be very spiky, and have a $C^1$ discontinuity at the origin. Although there is some debate as to whether this is unrealistic~\citep{VFW} or not~\citep{KHCB},~\citet{CFL} show that this, when combined with an assumption of asymptotic flatness, can be critical in deciding if SNIa contain enough information to test the Copernican principle: the distance modulus must differ significantly from the $\LCDM$ one.

In this paper we investigate this issue using the latest SNIa dataset, which includes SNIa at very low redshift ($z_\mathrm{min}\sim0.015$). The void profile can be made sufficiently differentiable by shaving off the spike in the void profile at any radius. Below $z = z_\mathrm{min}$, objects are not entirely in the Hubble flow and thus measurements are dominated by peculiar velocities, meaning that it is not possible to constrain cosmological models using the distance-redshift relation in this range. We introduce a simple void parameterisation which can reproduce the $\LCDM$ distance modulus to sub-percent accuracy, and which has a continuous parameterisation from a steep and spiky void mimicking $\LCDM$ to a smooth one. Thus we demonstrate {\it stricto sensu} that we can't differentiate between $\LCDM$ and LTB voids using SNIa alone. However, the general gist of~\citet{CFL} is that the void must be very steep to mimic $\LCDM$ distances closely, and that this is unnatural. By fitting our parameterisation to the new data directly we can see if a steep void is preferred over a gentle Gaussian profile. If a steep void is preferred by the data, then, this would lend weight to $\LCDM$. However, to fully break the degeneracy between $\LCDM$ and these void models another observable is required.

As suggested by the curvature test above, a good choice is $H(z)$. We use the ages of passively evolving galaxies to do this~\citep{Simon05}. This probe of the expansion rate of the Universe is still relatively new, and thus the data currently says little of significance. However, it has the benefit of being a relatively model independent method to reconstruct $H(z)$. In contrast to other works on this subject we don't use tests like the BAO  and the CMB. The reasoning for this is mainly that they are perturbative tests of the models and so test how perturbations evolve; this is particularly important for the BAO and small-$l$ CMB. The theory for this has not yet been worked out so we can't say whether results we would obtain have any meaning. Although the `background' part of the effect in LTB may be taken into account using the two Hubble rates in LTB~\citep{gbh1,gbh2,zibinvoid,bolejko}, it has been assumed in previous works that the perturbations evolve as they do in FLRW~-- without any scale dependence in their late-time evolution~-- and so a comoving sphere at last scattering remains so at late times, modulo the distortion from the different Hubble rates in the radial and angular directions. However, background curvature enters the Bardeen equation for the gravitational potential, and this can vary significantly over a sphere of radius 150 Mpc in void models; this will add an important additional distortion which is not yet known, and may affect the BAO significantly.

Furthermore, in LTB models, perturbations are complicated~\citep{pertpaper} because density perturbations couple to vector and tensor d.o.f.; how important this is in dissipating density fluctuations is also not known. Finally, we don't consider CMB constraints partly for the reasons discussed for the BAO: the large-scale CMB has not been calculated. The small-scale CMB may be estimated, however, and intriguingly seems to favour a non-zero bang-time function and asymptotic curvature~\citep{CF}. Though it is not yet clear how degenerate this result might be with the primordial power spectrum, which might be important because we don't have an inflationary model for void models, this is a very interesting result. Here, however, we are really mainly concerned with what the local data tells us about the shape of the void, assuming that these voids can evolve out of perturbed FLRW, and we don't consider CMB constraints further.

The paper is organized as follows. In \S\ref{sec:voids}, we describe the various void models based on the Lema\^itre-Tolman-Bondi metric and the kinematical quantities associated with them. Then in \S\ref{sec:data}, we discuss the cosmological data that we have used to constrain the models presented in \S\ref{sec:voids}. \S\ref{sec:analysis} is devoted to the data analysis itself, and the interpretation of the results; in particular, we discuss the ability of data to constrain the smoothness and the size of the void, and a series of non-concordance diagnoses. Finally, we conclude in \S\ref{sec:conc} with a summary of the main results and a sketch of future developments.

\section{Voids}
\label{sec:voids}

\subsection{Lema\^itre-Tolman-Bondi models}

We model the observable universe as an inhomogeneous void centered around us via the spherically symmetric LTB model with metric
\ba
\label{LTBmetric2}
ds^2 = -dt^2 + \frac{\apar^2(t,r)}{1-\kapr}dr^2 + \aperp^2(t,r)r^2d\Omega^2\,,
\ea
where the radial ($\apar$) and angular ($\aperp$) scale factors are related by
\ba\label{apar}
\apar \equiv (\aperp r)^{\prime}
\ea
and a prime denotes partial derivative with respect to coordinate distance $r$. The curvature $\kappa=\kappa(r)$ is not constant but is instead a free function. We choose coordinates such that the angular scale factor is constant and $a_\perp(t_0,r)=1$.
From these two scale factors we define two Hubble rates:
\ba\label{H}
\Hperp= \Hperp(t,r) \equiv \frac{\dot a_\perp}{\aperp},~~~~\Hpar=\Hpar(t,r) \equiv \frac{\dot a_{\|}}{\apar}
\ea
where an over-dot denotes partial differentiation with respect to
$t$. We denote their values today by $\Hperpo= \Hperpo(r)=H_\perp(t_0,r)$ etc.  The analogue of the Friedmann equation in this space-time is
then given by
\ba
\label{H}
H_{\perp}^2 = \frac{M}{a_{\perp}^3}-\frac{\kappa}{a_{\perp}^2},
\ea
where $M=M(r)$ is another free function of $r$, and the locally measured energy density is
\ba
8 \pi G \rho(t,r) = \frac{(M r^3)_{,r}}{a_{\parallel}a_{\perp}^2 r^2},
\ea
which obeys the conservation equation
\ba
\dot{\rho}+ (2 H_{\perp}+H_{\parallel}) \rho =0.
\ea
The acceleration equations in the transverse and radial directions are
\ba
\frac{\ddot{a}_{\perp}}{a_{\perp}} = -\frac{M}{2 a_{\perp}^3}
~~~~~\text{and}~~~~
\frac{\ddot{a}_{\parallel}}{a_{\parallel}} = -4 \pi G \rho +\frac{M}{a_{\perp}^3}.
\ea
We introduce dimensionless density parameters for the CDM and curvature, by analogy with the FLRW models:
\ba
\omegko(r)&=&-\frac{\kappa}{\Hperpo^2}, \nonumber\\
\omegmo(r)&=&\frac{M}{\Hperpo^2 } ,
\ea
using which, the Friedmann equation takes on its familiar form:
\ba
\frac{\Hperp^2}{\Hperpo^2}=\omegmo\aperp^{-3} + \omegko \aperp^{-2},
\ea
so $\omegmo(r)+ \omegko(r)=1$. Integrating the Friedmann equation from the time of the big bang $t_B = t_B(r)$ to some later time $t$ yields the age of the universe at a given $(t,r)$:
\ba
\label{solA}
\tau(t,r)= t - t_B = \frac{1}{\Hperpo(r)}\int^{\aperp (t,r)}_{0}\frac{d x}{\sqrt{\omegmo (r)x^{-1} + \omegko (r)}}\,.
\ea
(This integral can be given in closed form, but it's rather ridiculous.)
We now have two free functional d.o.f.: $\omegmo(r)$ and $t_B(r)$, which can be specified as we wish. However, if the bang time function is not constant this represents a decaying mode~\citep{silk,zibin}; consequently we set $t_B=0$ throughout, which means that our model evolves from FLRW. As a result, the age of the universe $\tau$ is then a constant, and equal to the time today $t_0$. So, by solving (\ref{solA}) for $\Hperpo (r)$, we have that:\\
\ba
\Hperpo(r)=\left\{
\begin{array}{cr}
 \displaystyle \frac{- \sqrt{-\omegko }+\omegmo\sin^{-1}\sqrt{-\frac{\omegko}{\omegmo}} } {t_0\left(-\omegko\right)^{3/2}}\,   &  \omegko<0 \\[4mm]
\displaystyle  \frac{2}{3t_0}   & \omegko=0   \\[2mm]
\displaystyle  \frac{\sqrt{\omegko } - \omegmo\sinh^{-1}\sqrt{\frac{\omegko}{\omegmo}} } {t_0\omegko^{3/2}}   &  \omegko>0
\end{array}
\right.
\ea
When we introduce the Hubble constant below, we shall use $H_0=\Hperpo(r=0)$, which fixes $t_0$ in terms of $H_0$, $\Omega_m(r=0)$ and $\Omega_k(r=0)$.

\subsection{Distance modulus and observables}
In LTB models, there are several approaches to finding observables such as distances as a function of redshift. We refer to~\citet{Enqvist} for details of the approach we use here. On the past light cone a central observer may write the $t,r$ coordinates as functions of $z$. These functions are determined by the system of differential equations
\ba
\label{dtdz}
\frac{dt}{dz} &=& -\frac{1}{(1+z)\Hpar}\,, \\
\label{drdz}
\frac{dr}{dz} &=& \frac{\sqrt{1-\kappa r^2}}{(1+z)\apar\Hpar},
\ea
where $H_\|(t,r)=H_\|(t(z),r(z))=H_\|(z)$, etc. The area distance is given by
\ba
d_A(z)=a_\perp(t(z),r(z)) r(z)
\ea
and the luminosity distance is, as usual $d_L(z)=(1+z)^2 d_A(z)$. Other observables can be calculated from these relations; in particular, the distance modulus is given by
\ba
\mu(z)=m-M=5\log_{10}\left[\frac{d_L(z)}{1\,\text{Mpc}}\right]+25,
\ea
where $m$ is the apparent magnitude of a source with intrinsic magnitude $M$.
The relative ages of galaxies
may be calculated from Eq.~(\ref{dtdz}) to give $H_\|(z)$.

The algorithm for calculating functions of $z$ is as follows:
\begin{enumerate}
\item Choose a profile $\Omega_m(r)$ today; this gives $\Omega_k(r)$ and $\Hperpo(r)$.
\item At any spacetime point $(t,r)$ we have $a_\perp(t,r)$ by inverting Eq.~(\ref{solA}).
\item We may then calculate $a_\|(t,r)$, and hence $H_\|(t,r)$, from Eq.~(\ref{apar}).
\item Integrate Eqs.~(\ref{dtdz}) and (\ref{drdz}) from $z=0$ with initial conditions $t(z=0)=t_0$ and $r(z=0)=0$ to give the parametric equations $t(z), r(z)$.
\item Calculate all functions as a function of $z$.
\end{enumerate}

\subsection{Void Profiles}
For our main model we introduce a profile which is capable of reproducing the $\LCDM$ distance modulus to high accuracy, as well as being able to control the smoothness at the centre of the void via the parameter $\nu$. The parameterization is given by (for $\nu>0$ and $\nu\neq1$)
\ba
\label{Eq:Profile1}
\omegmo(r)= \omegmin+\frac{\omegmout-\omegmin}{\nu-1}\left[\nu\tanh\frac{r}{\sigma}-\tanh\frac{r\nu}{\sigma}\right],\\
\hfil\color{red}(\text{Model \# 1})\nonumber
\ea
where $\omegmin$ and $\omegmout$ are the value of $\omegmo$ at the centre of the void at infinity, respectively. The parameter $\sigma$ characterizes the size of the void, but a more useful quantity is the full width at half-maximum (FWHM), calculated by solving
\ba
\omegmo\left(\frac{1}{2}\, r_{_{\textnormal{FWHM}}}\right) = \frac{\left(\omegmout + \omegmin \right)}{2}\,,
\ea
for $r_{_{\textnormal{FWHM}}}$ numerically. (We assume $\omegmo(r)$ is even about $r=0$ so that $\omegmo(-r)\equiv\omegmo(r)$.)

\begin{figure}
\begin{center}
\includegraphics[width=0.49\textwidth ]{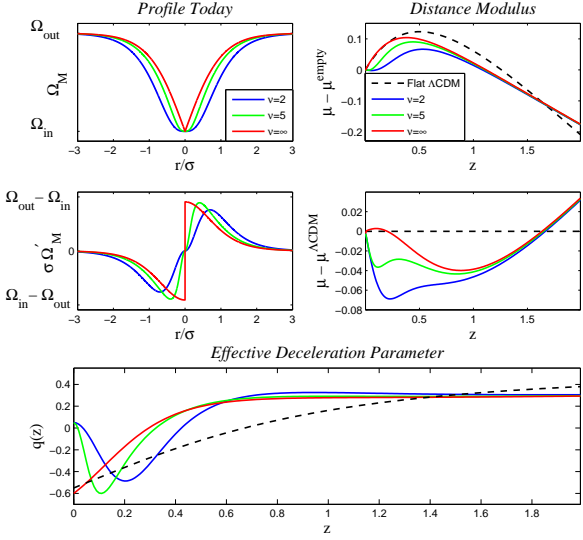}
\caption{Void \#1 shown for different values of $\nu$. For $\nu\rightarrow\infty$ we have a sharp, non-differential void which can mimic the $\LCDM$ distance modulus at low $z$ (middle right). (The effective deceleration parameter, shown bottom, is defined later.) Only when the profile has discontinuous derivative (middle left) can the deceleration parameter be negative at $z=0$.}
\label{fig:illustrating_nu}
\end{center}
\end{figure}

The parameter $\nu$ is chosen so that when $\nu$ is finite this function is $C^2$ at the origin (its first and second derivatives are well-defined and equal to zero at $r=0$). In the limit $\nu\to\infty$ we have
\ba
\omegmo(r)\to \omegmin+({\omegmout-\omegmin})\tanh\frac{r}{\sigma},
\ea
which can give an extremely good fit to the $\LCDM$ distance modulus~-- at the expense of not being differentiable at the origin.
The parameter $\nu$ gives us the power to control the sharpness of the void at the origin: the larger $\nu$ is, the steeper the void; the smaller $\nu$ is, the flatter the void is at the centre. Finally, note that the indeterminate form of $\Omega_m(r)$ at $\nu = 1$ is only due to the form of the parameterization: we can analytically continue the function (\ref{Eq:Profile1}) to include $\nu=1$:
\ba
\omegmo(r)= \omegmin+({\omegmout-\omegmin})\left[\tanh\frac{r}{\sigma}-\frac{r}{\sigma}\mathrm{sech}^2\,\frac{r}{\sigma}\right]\nonumber\\~~~\mbox{if}~~~\nu=1
\ea
In Fig.~\ref{fig:illustrating_nu}, we show this void for different values of $\nu$.

In addition to the parameterization given by Eq. (\ref{Eq:Profile1}), we also considered the following void profiles:
\ba
{\color{yellow}\text{\# 2}}:~~~\omegmo (r) &=& \omegmout - (\omegmout - \omegmin )\exp\left[-\left(\frac{r}{\sigma}\right)^2\right ]\,,
\nonumber\\
{\color{green}\text{\# 3}}:~~~\omegmo (r) &=& \omegmout - (\omegmout - \omegmin )\frac{\sigma^2}{\sigma^2 + r^2}\,, \nonumber\\
{\color{blue}\text{\# 4}}:~~~\omegmo (r) &=& \omegmout - (\omegmout - \omegmin )\frac{\sigma \sin\left( \frac{r}{\sigma} \right)}{r}\,, \nonumber\\
{\color{magenta}\text{\# 5}}:~~~\omegmo (r) &=& \omegmout - (\omegmout - \omegmin )\frac{\sigma^2 \sin^2\left( \frac{r}{\sigma} \right)}{r^2}\,.\nonumber
\ea

\begin{figure}
\begin{center}
\includegraphics[width=0.45\textwidth ]{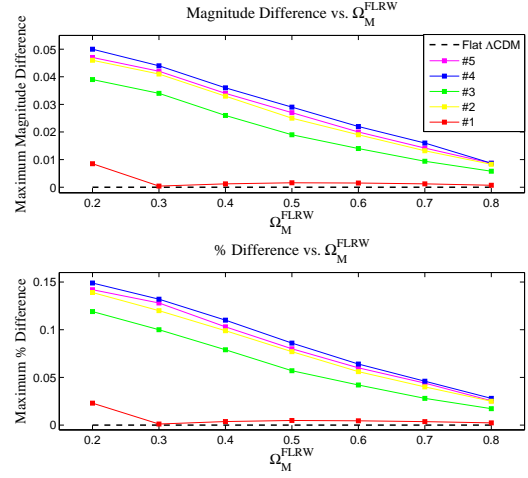}
\caption{For a given flat $\LCDM$ model we find the ability of each void model to reproduce the same $\mu(z)$. Here we show the largest differences as a function of the matter content of the FLRW model for the best-fitting void model in each case. }
\label{fig:percent-error1}
\end{center}
\end{figure}
\begin{figure}
\begin{center}
\includegraphics[width=0.45\textwidth , clip ]{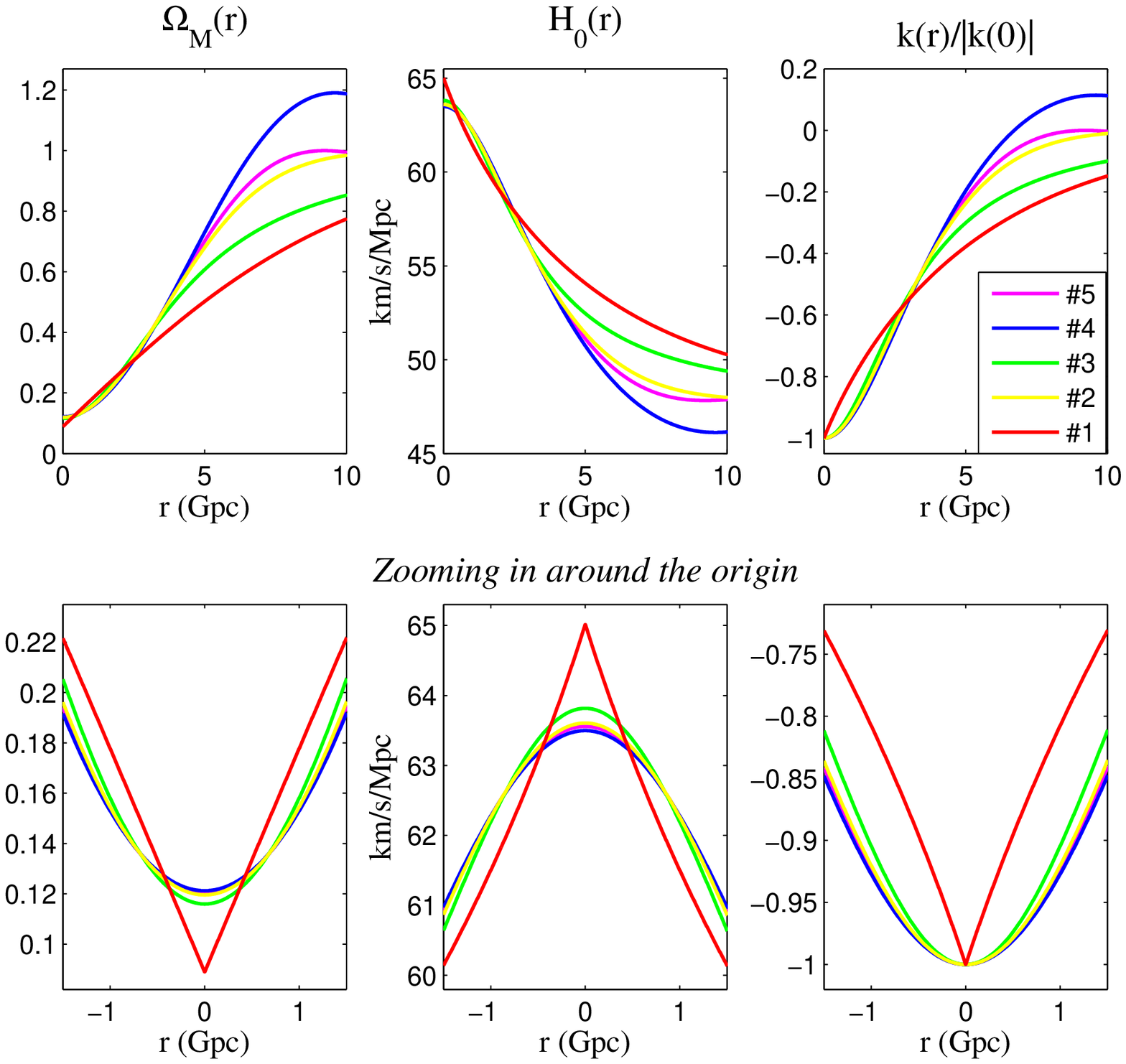}\\
\includegraphics[width=0.45\textwidth ,clip ]{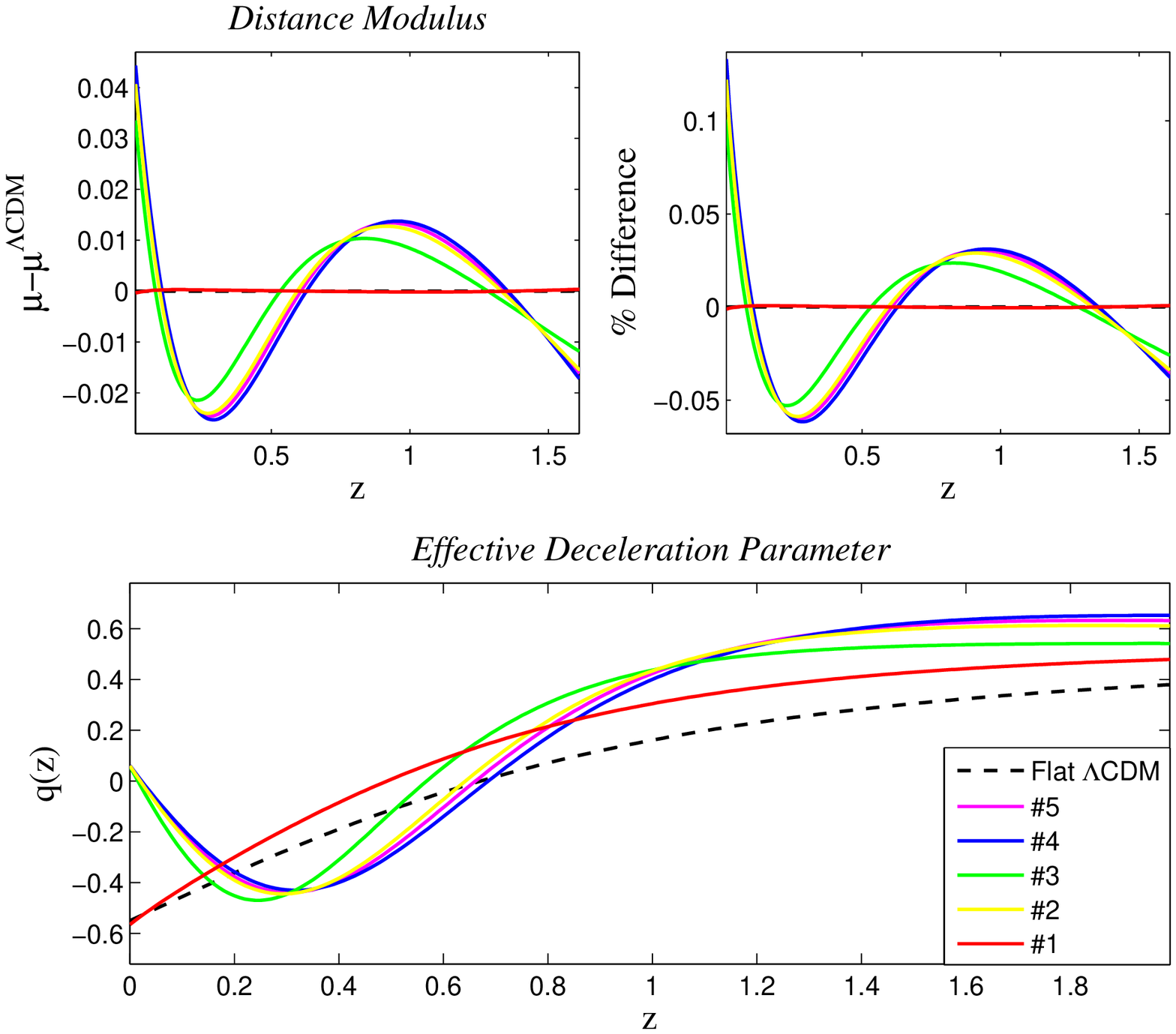}
\caption{Present day radial profiles (top 6 panels) and corresponding difference in distance moduli for each void and that of $\LCDM$ (third row from top), as well as the resulting effective deceleration parameter as a function of redshift (bottom), after fitting each void with $\omegmout = 1$ to $\LCDM$ with $\Omega_m^\text{FLRW}=0.3$.}
\label{fig:profiles}
\end{center}
\end{figure}
Typically we will fix $\omegmout=1$, so that the spacetime is asymptotically flat, in keeping with generic predictions from inflation. However, early-universe models which might produce a void of the kind we are considering have not been explored (although see~\citet{linde}). On the other hand, because we have set the bang time to $t_B = 0$, the models we consider evolve from a perturbed FLRW model at early times (the void has $|\delta\rho/\rho|\sim10^{-3}$ at last scattering), so this may conceivably be natural.

It is known that there exist LTB models which can give the FLRW distance modulus exactly for a central observer~\citep{MHE,CBKH,yoo}. Profile \#1 is a void parameterisation which can accurately mimic $\LCDM$ to high precision.
If we assume  $\omegmout=1$ and perform a least-squares fit of the void $\mu(z)$ to $\LCDM$ models for $0<z<1.6$, then we find that all of the profiles given can reproduce the distance modulus of $\LCDM$ to sub-percent accuracy~-- see Fig.~\ref{fig:percent-error1}. Model \#1 can produce a $\LCDM$ distance modulus to very high accuracy, which requires $\nu=\infty$ (a spiky void). The corresponding radial profiles for our 5 different best-fitting to $\LCDM$ void models, as well as their distance moduli and effective deceleration parameter (defined below by Eq.~(\ref{qeff})) are shown in Fig.~\ref{fig:profiles}. For Model \#1, even though the distance modulus is effectively the same as that of FLRW, the deceleration parameter is noticeably different; this is because the Hubble rates are different.

Finally, note that with $\omegmout=1$, voids 2-5 have 3 parameters (including $H_0$), which is the same as a curved $\LCDM$ model, while \#1 has 4. Note also that the parameter $\sigma$ has dimensions of length, an issue we will return to.

\subsection{Physical length scales and the distance modulus}\label{sec:H0}

In LTB models, subtleties arise concerning the relation between $H_{0}$ and the magnitude-redshift relation. In the FLRW case, since $H_0$ is just a magnitude offset related to the intrinsic (absolute) luminosity of a SNIa, it is usually removed from the analysis by marginalizing over it. The intrinsic absolute magnitude of SNIa's is poorly constrained, and since this value and the value of $H_0$ are degenerate in $\LCDM$, the Hubble constant is poorly measured by SNIa.

In an LTB model we have two length scales which are independent: the Hubble length, $H_0^{-1}$, associated with the expansion time, and the void scale depending on the physical `size' of the underdensity. In the models considered here, this may be characterized by the parameter $\sigma$ or the FWHM. The ratio of these two scales enters as a dimensionless number, and this must therefore have a physical significance. Thus, when we change $H_0$ keeping the void scale fixed this will be reflected in the distance modulus in a non-trivial way, namely, if one rescales the distance by $r\mapsto \hat r/H_0$, in the void profiles this becomes $r/\sigma=\hat r/\sigma H_0$.

\begin{figure}
	\centering
	\includegraphics[width=0.45\textwidth ]{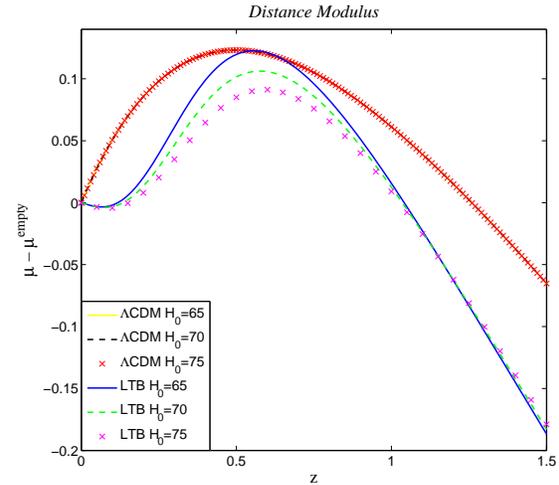}
	\caption{Plots of $\mu - \mu^{\textnormal{empty}}$ for FLRW and LTB models for different choices of $H_0$. This illustrates that, in the LTB case, $H_0$ not only affects the vertical displacement of $\mu(z)$, but also its shape. This means that our obtained void parameters are partially dependent on this value, and thus a standard normalisation cannot be applied.}
	\label{fig:eg_H_0}
\end{figure}

As a result, higher (lower) values of $H_0$ not only shifts the distance modulus curves down (up), but it also affects the overall shape too.  This is shown in Fig.~\ref{fig:eg_H_0}, where we can see that the shape of the $\mu(z)$ function changes with $H_0$. If $\sigma H_0$ were held fixed when $H_0$ is changed (for example by fixing $\sigma$ in units of $h^{-1}$Mpc rather than Mpc, say), this would compensate for this effect and then $H_0$ would be a pure normalisation of the distances. However, linking the two independent scales in this way is rather restrictive.

This is an important issue, and, as far as we are aware, has not been previously considered. This means that the shape and size of our best fit void model is dependent on the value of $H_0$ we obtain when we fit the voids to data. With SNIa being able to poorly constrain the value of $H_0$, this leads to an additional uncertainty in the best-fitting parameters obtained. The best-fitting value of $H_0$ as indicated by the latest supernova dataset is clearly not in agreement with other measurements, such as that found by the HST Key Project using Cepheid variables. However, we estimate that the additional uncertainty in void models is below the $5\%$ level, and as such does not play a significant role in the current error budget. Future studies of inhomogeneous models, be it globally LTB models as in our case, or mass-compensated ones in other cases (see e.g. ~\citet{KH},~\citet{notari}), will need to consider the effect of $H_0$ on their results, possibly by fitting the supernova light-curves simultaneously with the model.

\section{Cosmological Data}
\label{sec:data}

In this paper, we confront each of the 5 void models we introduced in the last section with the largest sample of SNIa to date; the Constitution dataset consisting of 397 SNIa ~\citep{Hicken09b}, as well as the $H(z)$ data ~\citep{Simon05}, consisting of 10 points when we include the HST Key Project value of the Hubble constant, $H_0 = 72 \pm 8$ \hu ~\citep{freedman}.

The Constitution dataset of SNIa ~\citep{Hicken09b} is comprised of a large sample of nearby ($z<0.08$) objects ~\citep{Hicken09a} combined with the Union dataset ~\citep{Kowalski08} which covers a redshift range up to $z=1.85$. This sample comprises 397 spectroscopically confirmed SNIa, making it the largest publicly available and uniformly analysed SNIa dataset to date. An additional intrinsic dispersion of $\sigma_{\mu}$ = 0.12 is added to the errors of each supernova to better estimate the error due to the uncertainty in absolute magnitude of these events.

In \S\ref{sec:analysis} we consider several other supernova datasets. These samples do not include the large number of SNIa's at low redshift that comprise a large proportion of the Constitution dataset. The Davis et al. dataset \citep{Davis07} combines SNIa's discovered from the ESSENCE (Equation of State: SupErNovae trace Cosmic Expansion) survey with those from the High-Z release to produce a sample of 192 objects. The Union dataset \citep{Kowalski08} comprises 307 objects that have been combined from other publicly released datasets, whilst the ConstitutionT sample \citep{truncated} uses the Constitution dataset described above, with 34 SNIa removed in order to make the sample be in better agreement with other SNIa datasets (not to mention the $\LCDM$ model!).

The expansion rate of the Universe, $H(z)$ has been constrained using the differential ages of passively evolving galaxies as determined by fitting SED templates to their spectra. Using data from the Gemini Deep Survey ~\citep{Abraham04} and a sample of field early-type galaxies ~\citep{Treu99, Treu01, Treu02} along with two radio galaxies ~\citep{Dunlop96, Spinrad97, Nolan03}, ~\citet{Simon05} were able to constrain the evolution of $H(z)$ for $0.1 < z < 1.8$. In this analysis we include the measurement of the local value of $H(z)$ as determined by ~\citet{freedman} to make 10 data points, but note that a recent, more accurate measurement of $H_0$ \citep{riess09} and additional $H(z)$ points \citep{jimenez09} have since been determined, and would place somewhat tighter contraints on the models considered in this work.

Thus this analysis uses two independent probes of the expansion history of the Universe. Other work, such as~\citet{zibin}, combine supernova datasets with other probes, such as BAOs and CMB measurements; however, these are not considered in this analysis, as explained in \S\ref{sec:intro}, since the growth of perturbations has not been properly explored in LTB cosmologies and thus approximations to $\LCDM$ relations may not be valid.

\section{Data Analysis}
\label{sec:analysis}

\subsection{Overview of Numerical and Statistical Method}
We made use of the publically available easyLTB Fortran90 code provided by \citet{gbh1} to run through more than a million different parameter values, simultaneously computing $\mu(z)$, $H(z)$ and the $\chi^2$ statistic for each model with the intent of finding the best-fitting parameters for the voids considered.
The code is setup to compute $\chi^2$ for the SNIa ($\chi^2_{_{SNIa}}$), BAO ($\chi^2_{_{BAO}}$) and CMB data ($\chi^2_{_{CMB}}$), but we have left the testing of the LTB models against the last two for future work due to the fact that these quantities are affected by the growth of structure in the universe. On the other hand, we have added the Hubble rate data into our analysis, and we have used the longitudinal Hubble rate $\Hpar$ as our model prediction, since~\citep{Simon05} determined $H(z)$ via $dt/dz$ measurements, which in the LTB case (see Eq. (\ref{dtdz})) is related to $\Hpar$. The main reason for adding the $H(z)$ data into our analysis is because, as shown in \S\ref{sec:voids}, all of our void parameterizations can mimic $\mu^{\LCDM}$, making it impossible for us to distinguish between two such models \emph{if the data prefers $\LCDM$}; so using $H(z)$ in addition to the SNIa data (which we refer to as SNIa+H) allows us the further opportunity to find detectable differences between $\LCDM$ and LTB void models.
The easyLTB code is not set up to compute models in which $\omegmo(r) > 1$. For this analysis, we modified the code to allow $\omegmo(r)$ to take on any value, as required by our oscillating voids, although for our main investigation we have fixed $\omegmout$ to 1.

Regarding the fitting of our void models to data, we computed the $\chi^2$ statistic, and associated reduced $\chi^2$, $\chi^2_{\textrm{red}}$, to determine the goodness of fit of the model to the data. Cases where $\chi^2_{\textrm{red}} < 1$ indicate that the additional intrinsic dispersion added to the type Ia supernova error estimates is a conservative choice. On top of this, the confidence limits on each of the parameters was calculated using their likelihood distributions.

\subsection{Pointy or Smooth?}

As demonstrated earlier, void models that are pointy at the origin are capable of reproducing $\mu^{\LCDM}$ to arbitrary accuracy. However, what level of smoothness/sharpness at the origin do the SNIa data demand?

\begin{figure}
\centering
\epsfig{file=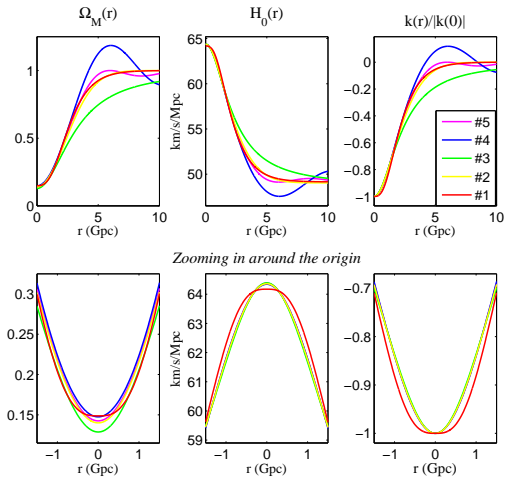,width=0.45\textwidth }
\caption{Best-fitting void profiles from SNIa data.}
\label{fig:FIT2_SN_DATA_Profiles}
\end{figure}

In Table \ref{table:othersndata}, we show the best-fitting model parameters of void \#1 when fitting to the samples described in \S\ref{sec:data}. The Davis et al. sample ~\citep{Davis07} (A), the Union sample ~\citep{Kowalski08} (B), the original Constitution sample ~\citep{Hicken09b} (C), the Constitution sample with intrinsic dispersion added (D, our principle SNIa dataset) and finally the ConstitutionT  sample ~\citep{truncated} (E) are considered.
\begin{table*}
\begin{minipage}{126mm}
\caption{Best-fitting parameters of void model \#1 from various SNIa data sets.}
\centering
\medskip
\begin{tabular}{|c|ccccccc|}
\hline
Data&$H_0$ &$\omegmin$&$\sigma$&$\log_{10}\nu$&$\chi^2_{SNIa}$ & $\chi^2_\textrm{red}$&FWHM \\ [0.2ex]
  Set $\#$ &    &    &Gpc    &                &      &						   &Gpc \\ [0.1ex]
\hline
A & 65.78&0.065& 9.26 & 2.95 & 195.38 & 1.03   & 10.19\\ [0.1ex]
B & 70.60&0.095 & 6.50&  5.00 & 310.68 & 1.02 & 7.14 \\ [0.1ex]
C &64.18 &0.140& 2.17& 8$\times 10^{-6}$ & 459.93 & 1.17 & 5.44 \\ [0.1ex]
D &64.17 &0.149& 2.05& 4$\times 10^{-6}$ & 308.82 & 0.78 & 5.13 \\ [0.1ex]
E &64.00&0.158 & 2.33& 3.4$\times 10^{-2}$ & 270.88 &0.75 & 5.52\\ [0.1ex]
\hline
\end{tabular}
\label{table:othersndata}
\end{minipage}
\end{table*}
\begin{figure}
\centering
\epsfig{file=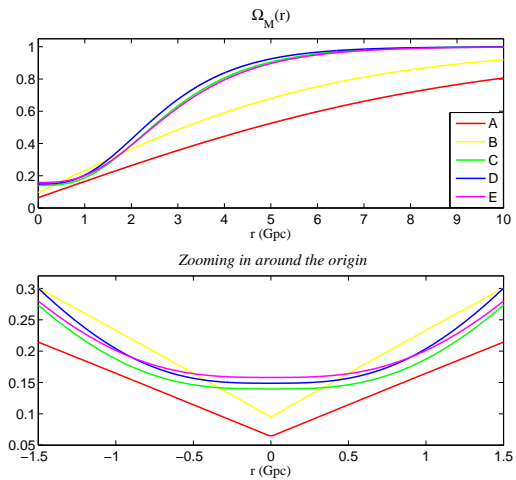,width=0.45\textwidth }
\caption{Best-fitting $\omegmo(r)$ profiles for void model \#1 from various SNIa datasets.}
\label{fig:FIT2SN_SpikeySmooth_Profiles}
\end{figure}
From these results we see that the best-fitting $\nu$ is different for different data sets. The reason is as follows: in the older data sets such as A and B, there are few to none low-redshift SNIa, so that $\nu$ is essentially meaningless, although, as shown in Table \ref{table:othersndata} and Fig. \ref{fig:FIT2SN_SpikeySmooth_Profiles}, the data sets given by A and B choose sharp voids naturally. On the other hand, with the abundant low-redshift SNIa in data sets C-E, $\nu$ can play an active role in determining the gradient of the profile around the origin, and, in particular, these data sets favour $\nu \sim 1$, corresponding to a void that is very smooth around the origin. This supports the claim by~\citet{CFL} that ``realistic'' voids should be smooth around the origin. Moreover, as Fig.~\ref{fig:FIT2_SN_DATA_Profiles} shows, void model \#1 turns out to be flatter around the origin than the Gaussian model (\#2).
Note that even with the next generation of SNIa measurements, a very low $z$ distinction will not be realistic, since below $z\sim0.015$ SNIa are dominated by peculiar velocities and are not properly in the Hubble flow~-- indeed, a cosmological model becomes a bit meaningless on such small scales.

\begin{table*}
\begin{minipage}{126mm}
\caption{Best-fitting void model parameters from SNIa data.}
\centering
\medskip
\begin{tabular}{|c|cccccc|}
\hline
Model&$H_0$ &$\omegmin$&$\sigma$&$\chi^2_{SNIa}$& $\chi^2_\textrm{red}$ &FWHM \\ [0.2ex]
 $\#$& \hu &      &Gpc     &             &						           &Gpc \\ [0.1ex]
\hline
$\color{yellow}\text{2}$& 64.36&0.140 & 3.29 & 308.83 &0.782 &5.47 \\ [0.1ex]
$\color{green}\text{3}$&64.39 &0.129& 3.21&  309.36 & 0.783 & 6.42 \\ [0.1ex]
$\color{blue}\text{4}$&64.34&0.148 & 1.35& 308.54 &0.781 & 5.77\\ [0.1ex]
$\color{magenta}\text{5}$& 64.35&0.143 &1.90 & 308.70 &0.782 & 5.29 \\ [0.1ex]
\hline
\end{tabular}
\label{table:bfsndata}
\end{minipage}
\end{table*}
In Table~\ref{table:bfsndata} we show the best-fitting parameters for void models \#2-5 after fitting them to the Constitution SNIa sample with intrinsic dispersion added (D). It is interesting to note that all of these models fit the SNIa data equally well compared to void model \#1~-- see Table~\ref{table:othersndata}. In addition, the FWHM of each void is roughly the same. Therefore, the simple void profiles such as the ones we have studied can all agree on the characteristic size that a void should be to fit the data, implying that one could effectively use any of these toy models in order provide a reasonable estimate of the physical size a typical void should be in order to fit the data. Comparing the $\chi^2$'s of the voids (model \#1 included) with that of the $\LCDM$ model in Table~\ref{table:bfLCDM} we obtain a similar (although slightly better) fit to the SNIa data as flat $\LCDM$.
\begin{table}
\caption{Best-fitting parameters for the flat $\LCDM$ model from SNIa and SNIa+H data. See the caveats in \S\ref{sec:H0} regarding what the value of $H_0$ means.}
\centering
\medskip
\begin{tabular}{|c|cccc|}
\hline
Data & $H_0$ & $\omegam$ & $\chi^2$ & $\chi^2_\textrm{red}$\\ [0.2ex]
\hline
SNIa   & 64.93  &  0.29 & 311.18 & 0.786 \\[0.1ex]
SNIa+H &  64.55  &  0.32 & 326.51 & 0.804 \\[0.1ex]
\hline
\end{tabular}
\label{table:bfLCDM}
\end{table}

\begin{figure}
\centering
\epsfig{file=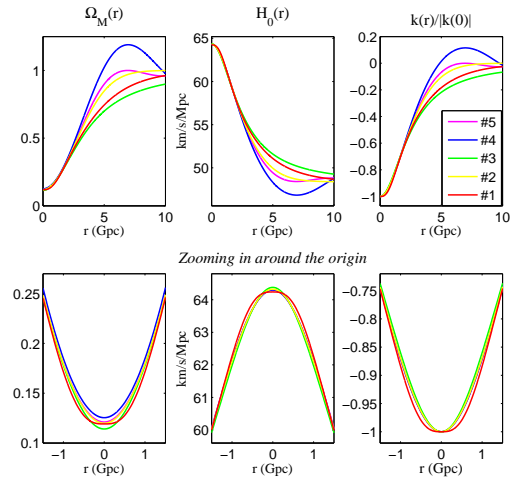,width=0.45\textwidth}
\caption{Best-fitting void profiles from SNIa+H data.}
\label{fig:FIT2_CMBND_DATA_Profiles}
\end{figure}

Given the difficulty of distinguishing LTB voids from $\LCDM$ using the SNIa data alone, we performed a combined analysis of the SNIa sample denoted D above with the $H(z)$ data. In Table~\ref{table:bfcmbnddata} we show the best-fitting parameters from SNIa+H data for each of our 5 void models, along with the corresponding $\chi^2$'s and FWHM. Notice again that the $\chi^2$, and $\chi^2_\textrm{red}$ values for the voids are not only comparable to that of $\LCDM$ (see Table~\ref{table:bfLCDM}), but are slightly lower. In Fig.~\ref{fig:FIT2_CMBND_DATA_Profiles} we show the resulting void profiles for the best-fitting to SNIa+H data. The sizes (FWHM) of the voids are slightly larger (by roughly 1 Gpc) in this case compared to that of the fit to SNIa case. This is a result of the fact that the fit to $H(z)$ by itself favours enormous voids (FWHM $\sim$ 65 Gpc!), and thus when combined to the SNIa data, we obtain bigger voids.
\begin{table*}
\begin{minipage}{126mm}
\caption{Best-fitting void model parameters from SNIa+H data.}
\centering
\medskip
\begin{tabular}{|c|ccccccc|}
\hline
Model \#&$H_0$ &$\omegmin$&$\sigma$&$\log_{10}\nu$&${\chi}^2$&$\displaystyle{\chi^2_\textrm{red}}$&FWHM \\ [3mm]
\hline
$\color{red}\text{1}$& 64.24&0.119&4.94 & 0.71 & 321.64 & 0.796  & 6.81\\ [0.1ex]
$\color{yellow}\text{2}$& 64.33&0.120 & 3.77&  $-$ & 321.74 &0.794  & 6.28 \\ [0.1ex]
$\color{green}\text{3}$& 64.38 &0.114& 3.59& $-$ & 321.57 & 0.794 & 7.17\\ [0.1ex]
$\color{blue}\text{4}$& 64.30&0.125 & 1.55&$-$ & 321.81 &0.795  & 6.65\\ [0.1ex]
$\color{magenta}\text{5}$& 64.35&0.121 &2.21 &$-$ & 321.86 &0.795 & 6.14 \\ [0.1ex]
\hline
\end{tabular}
\label{table:bfcmbnddata}
\end{minipage}
\end{table*}

\begin{figure}
\centering
\epsfig{file=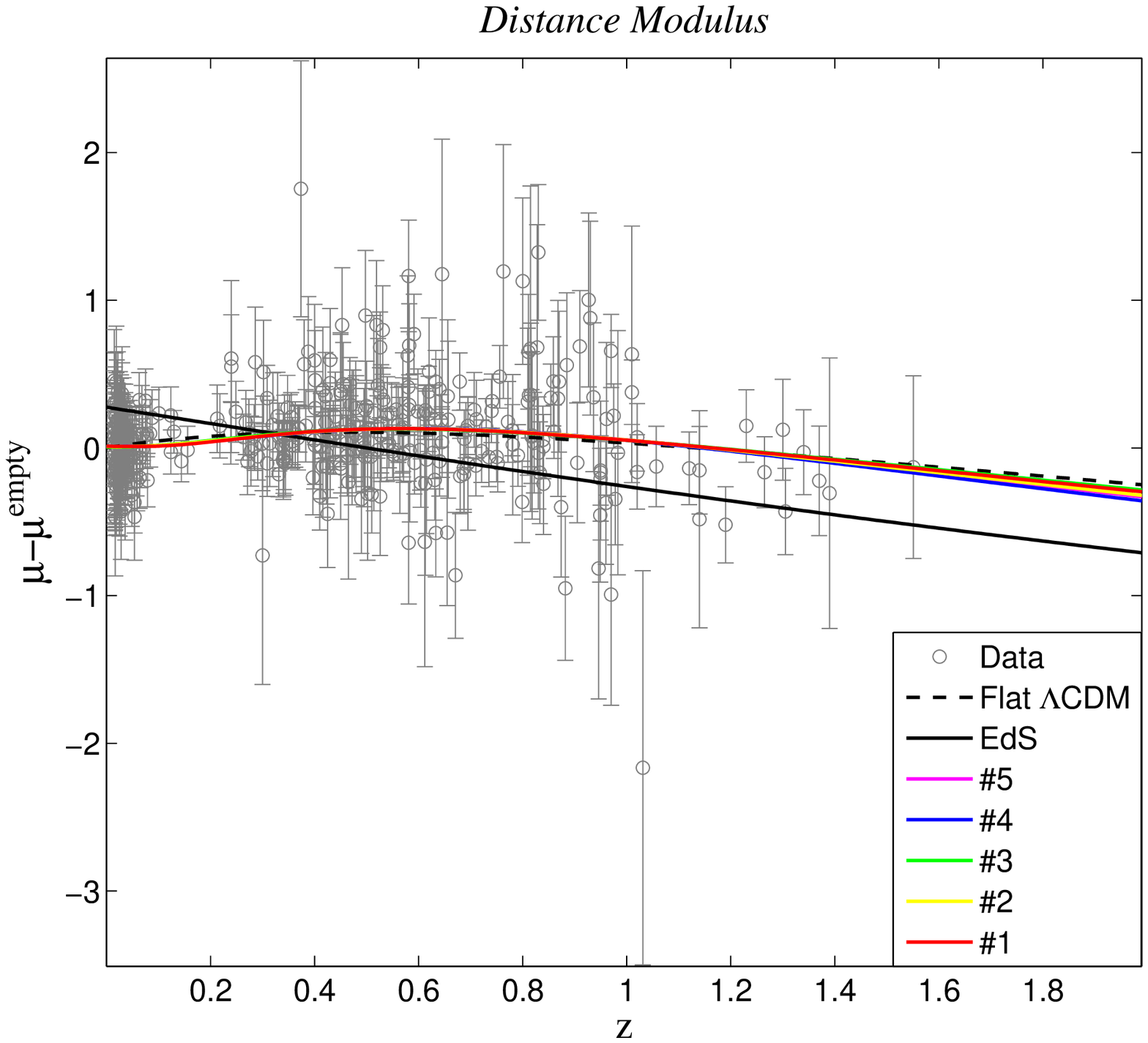,width=0.45\textwidth , clip} \\
\epsfig{file=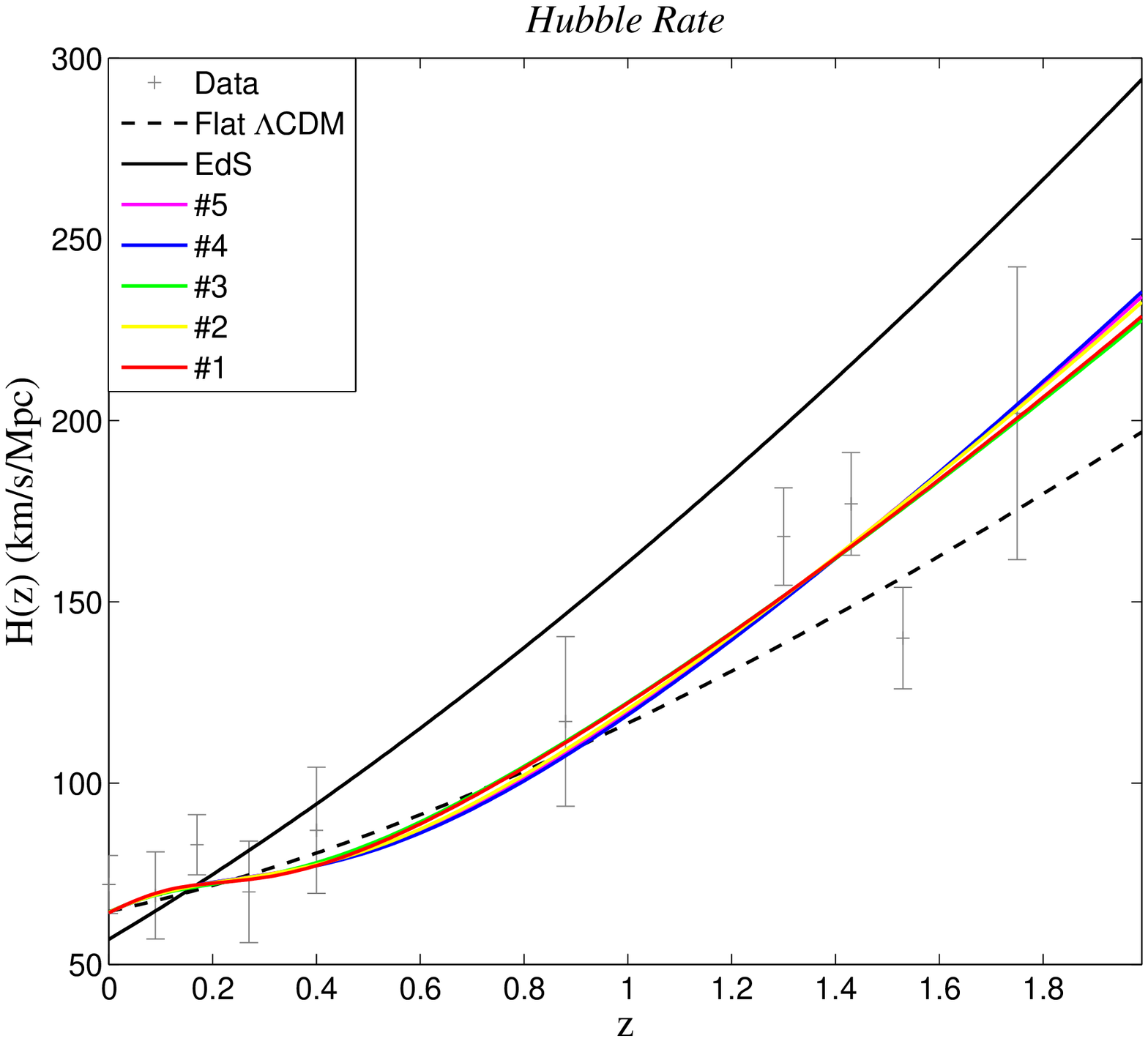,width=0.45\textwidth , clip}
\caption{Top panel: Best-fitting SNIa+H distance moduli, normalized to that of a Milne universe with $H_0$ given by that of the best-fitting $\LCDM$ model, for each void, along with that of the best-fitting $\LCDM$ and EdS models, as well as the Constitution SNIa data points (grey circles). Bottom panel: The best-fitting SNIa+H Hubble rates for each void model and that of $\LCDM$ and EdS, with the $H(z)$ data points overplotted (grey circles). Notice how well the void $H(z)$ curves follow the data points.}
\label{fig:bestfit_models_data}
\end{figure}

\begin{figure}
\centering
\begin{tabular}{c}
\epsfig{file=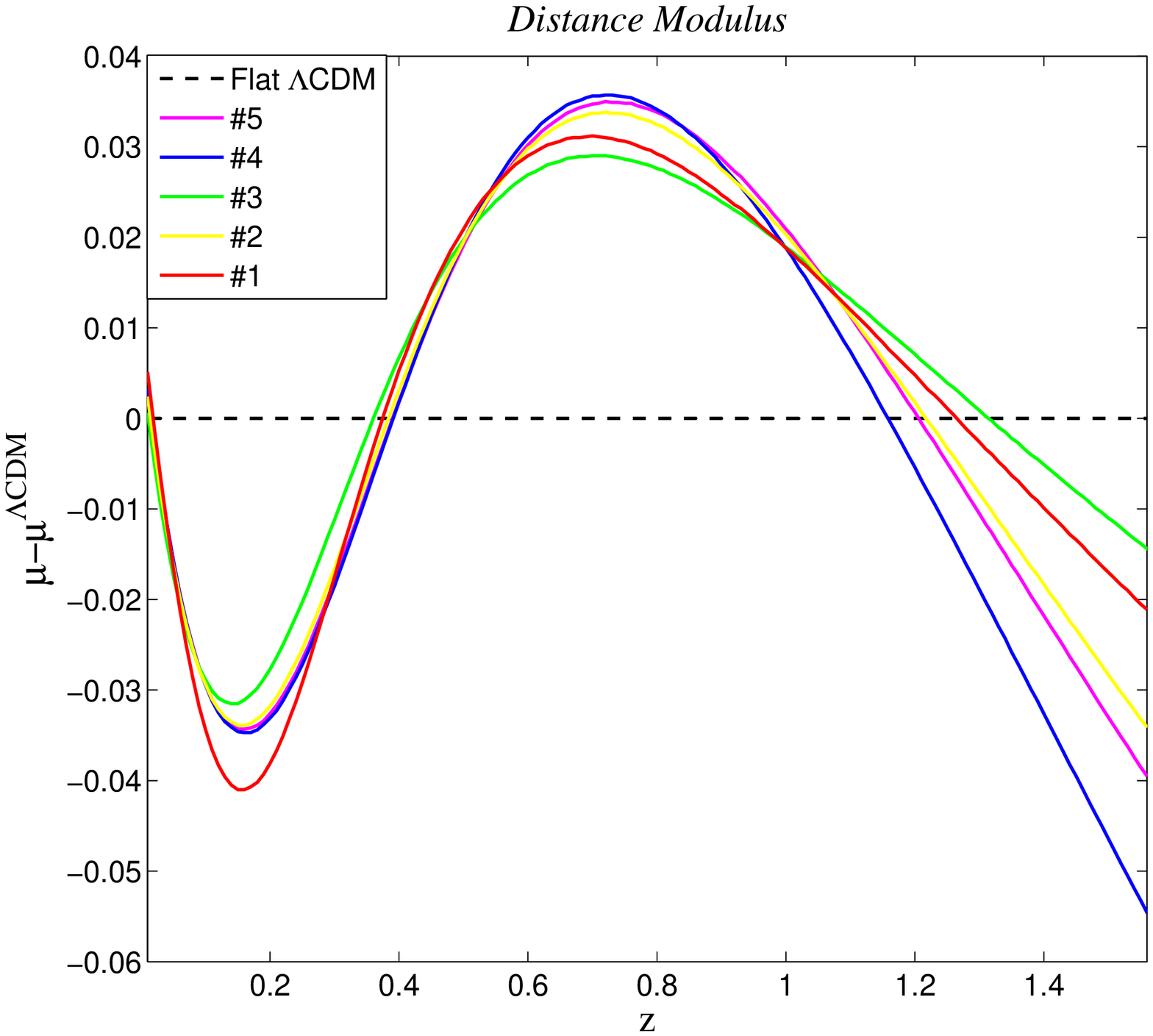,width=0.45\textwidth , clip}
\end{tabular}
\caption{Difference in distance moduli between the best-fitting SNIa+H voids and that of $\LCDM$. Up till a redshift $z\sim 1$, these void models are practically indistinguishable from that of $\LCDM$. There is no way to tell the difference between $\LCDM$ and our void models from SNIa alone, but as the bottom panel of Fig.~\ref{fig:bestfit_models_data} shows, $H(z)$ may play a key role in the future for testing for deviations from $\LCDM$.}
\label{fig:Mu_Minus_MuLCDM}
\end{figure}

In Fig.~\ref{fig:bestfit_models_data} we show the best-fitting to SNIa+H data residuals (top panel) and Hubble rate (bottom panel) for each of our 5 different void models, as well as that of $\LCDM$ and EdS. In each case, the respective data points are overplotted (grey circles), along with 1-$\sigma$ error bars. To illustrate the difficulty in distinguishing these voids from that of $\LCDM$ using SNIa data alone, in Fig.~\ref{fig:Mu_Minus_MuLCDM} we plot the magnitude difference between the two models after fitting our 5 different voids to the SNIa+H data.

\subsection{Likelihood Contours}
In this section, we explore the degeneracies between all the possible pairs of parameters by constructing joint-parameter likelihood contours, and also determine the likelihood distributions of each parameter from SNIa and SNIa+H constraints for void model \#1.

\begin{table*}
\begin{minipage}{126mm}
\caption{Marginalized best-fitting parameters at 95 per cent confidence from SNIa and SNIa+H constraints. Note the amusing error margins on the last two parameters individually! The error in the $\sigma-\nu$ projection are much more informative~-- see Fig.~(\ref{fig:likelihood_2D_CMBND})}
\centering
\medskip
\begin{tabular}{|c|cccc|}
\hline
Data & $H_0$ & $\omegmin$ & $\sigma$ (Gpc) & $\log_{10}\nu$\\ [0.2ex]
\hline
Priors & 58.00$-$74.00   & 0.00$-$0.24 & 0.00$-$30.00 & -1.30$-$3.70\\[0.1ex]
     &             &        &         &           \\ [-1.8ex]
SNIa   & $64.61^{+1.69}_{-1.39}$  &  $0.10^{+0.09}_{-0.07}$ & $6.72^{+17.68}_{-5.47}$ & $3.3^{+\infty}_{-3.97}$\\[0.2ex]
     &             &       &          &          \\ [-1.8ex]
SNIa+H & $64.72^{+1.58}_{-1.49}$	&  $0.10^{+0.07}_{-0.05}$ & $7.84^{+12.54}_{-6.27}$ & $3.5^{+\infty}_{-3.84}$\\[0.2ex]
\hline
\end{tabular}
\label{table:margbfsndata}
\end{minipage}
\end{table*}
Table \ref{table:margbfsndata} shows the marginalized best-fitting parameters for void model \#1 with 95 per cent confidence limits, from fitting to SNIa and SNIa+H data, where the priors are also given. We see again that the main difference between the parameter values in the SNIa and SNIa+H cases is the value of $\sigma$: the SNIa+H case yields a higher value for $\sigma$, which arises from the fact that, when fitting void models to $H(z)$ data only, extremely large ($\sigma\sim 20$) Gpc and empty ($\omegmin\sim0.01$) voids are the ones that give the minimum $\chi^2$, so when combining with the fit to SNIa results which has a lower value for $\sigma$, it is only natural that $\sigma$ is then larger in the fit to SNIa+H case than in the fit to SNIa case only. Notice in the column for the smoothness parameter $\nu$ (or log$_{10}\nu$ as shown), that the best-fitting values are roughly 3 orders of magnitude larger than that in the unmarginalized case (see Tables \ref{table:bfsndata} and \ref{table:bfcmbnddata}), indicative of sharp, spikier voids. Also note that there is no upper bound on $\nu$, only a lower one, but we shall return to this topic in a moment.

\begin{figure}
\centering
\begin{tabular}{cc}
\epsfig{file=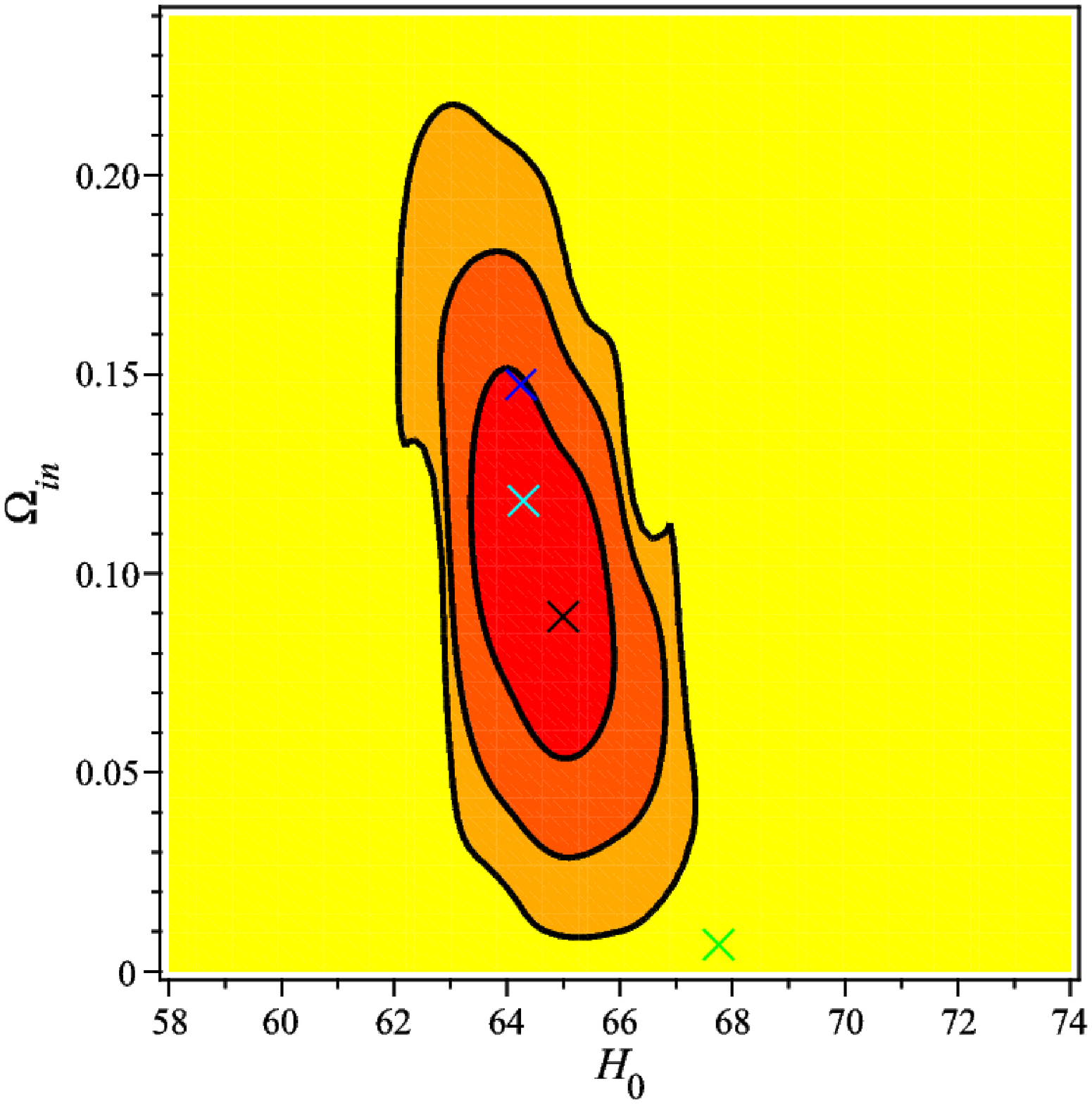,width=0.22\textwidth , clip} &
\epsfig{file=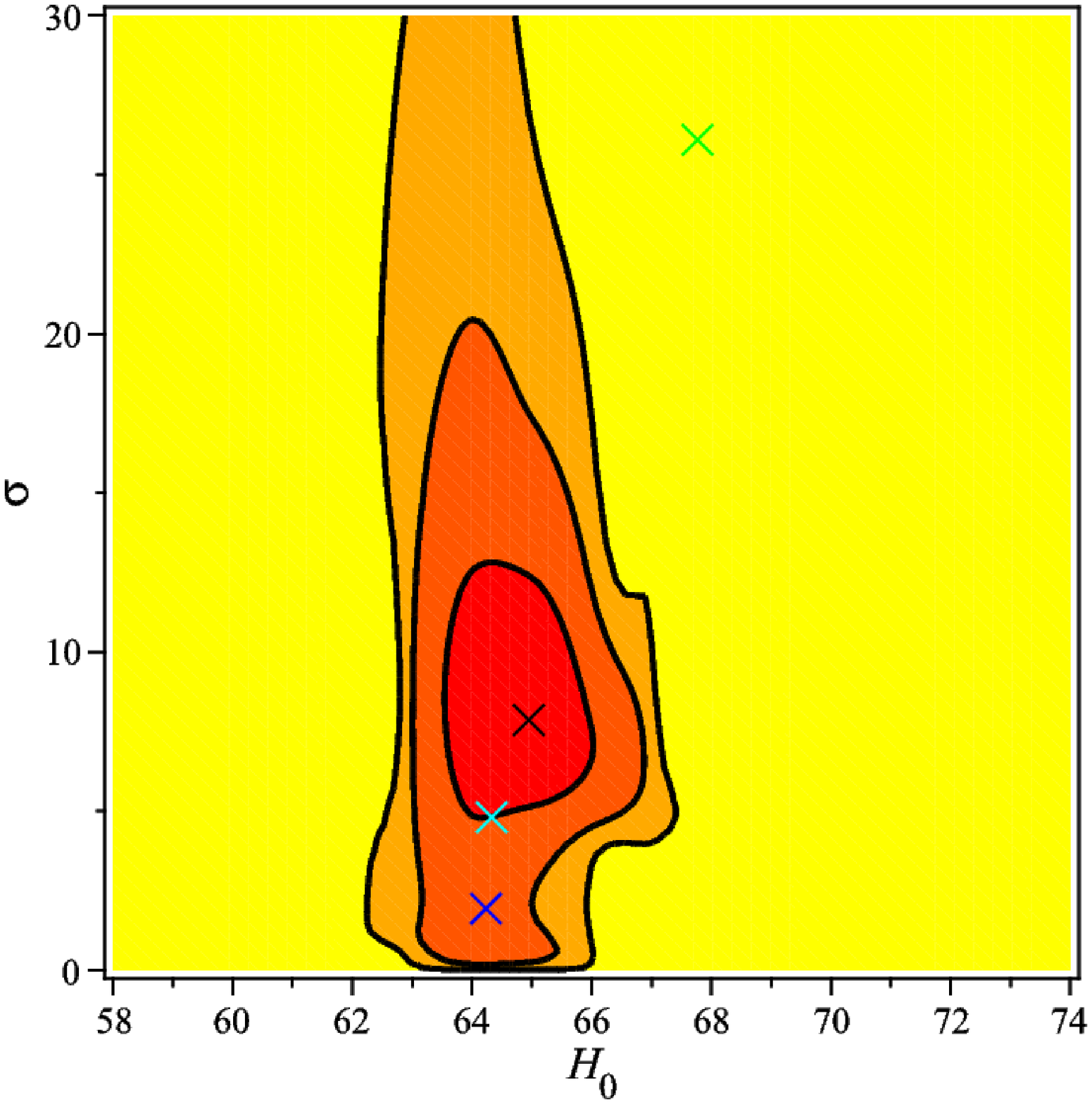,width=0.22\textwidth , clip}\\
\epsfig{file=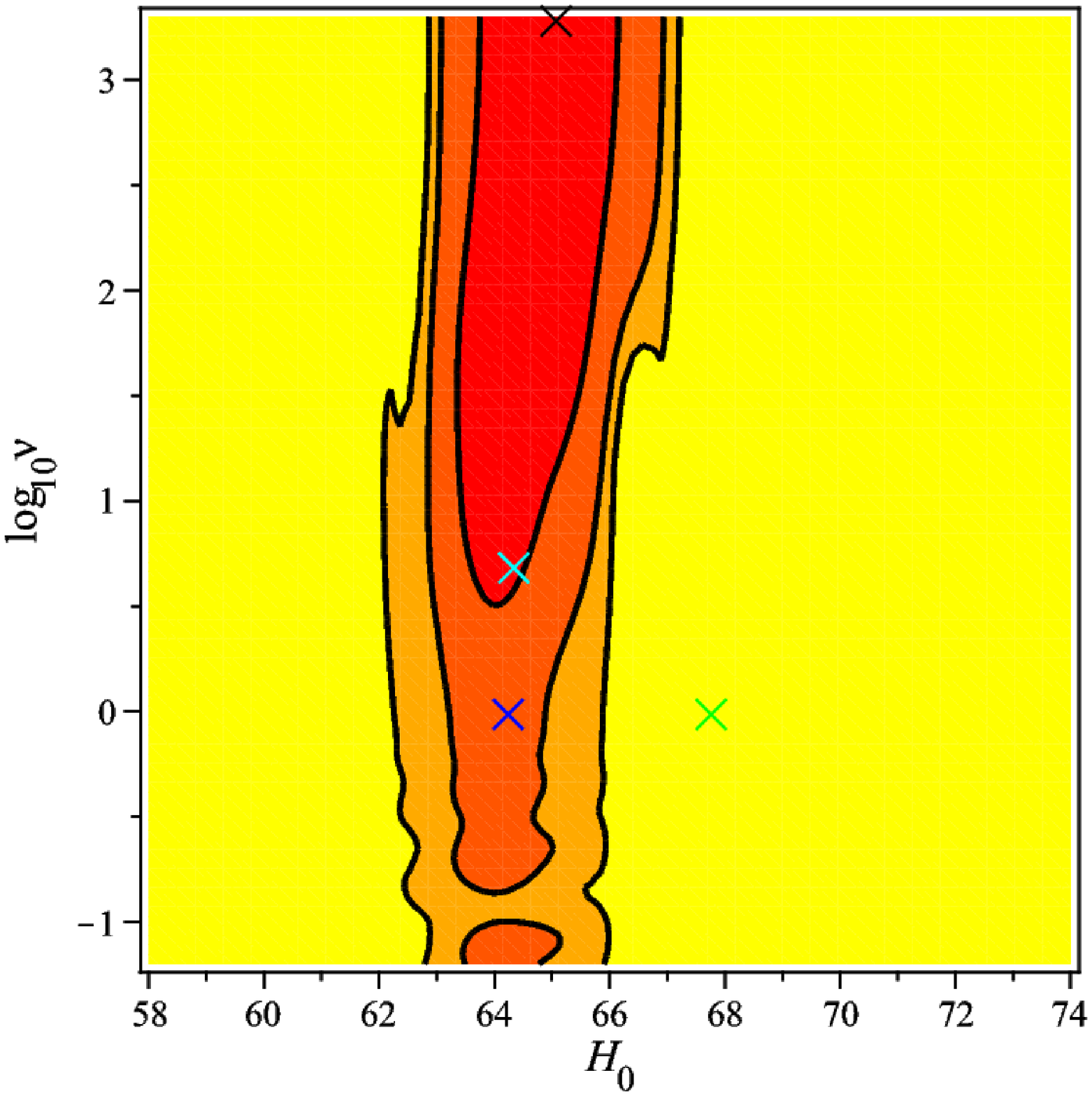,width=0.22\textwidth , clip} &
\epsfig{file=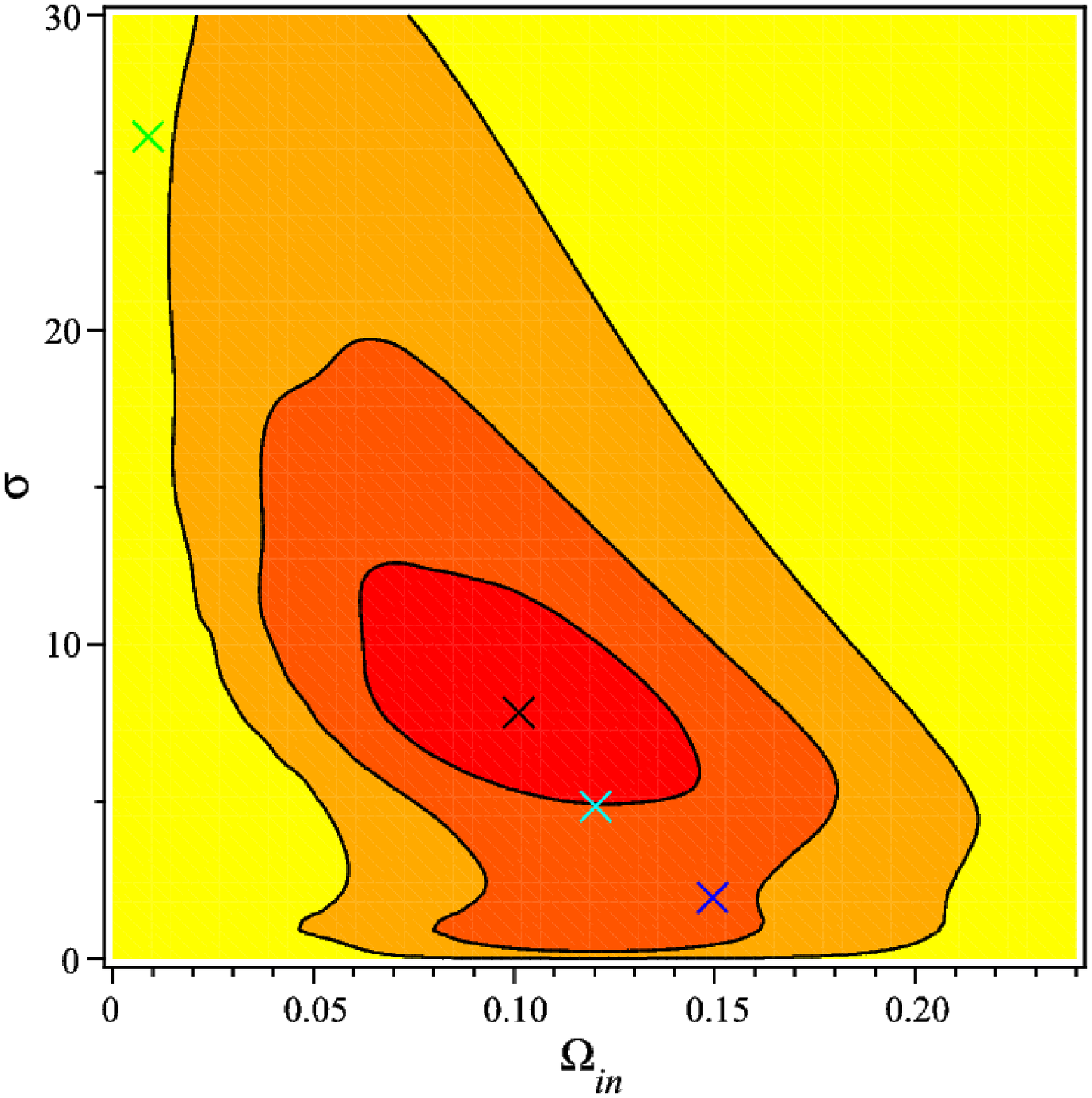,width=0.22\textwidth , clip}\\
\epsfig{file=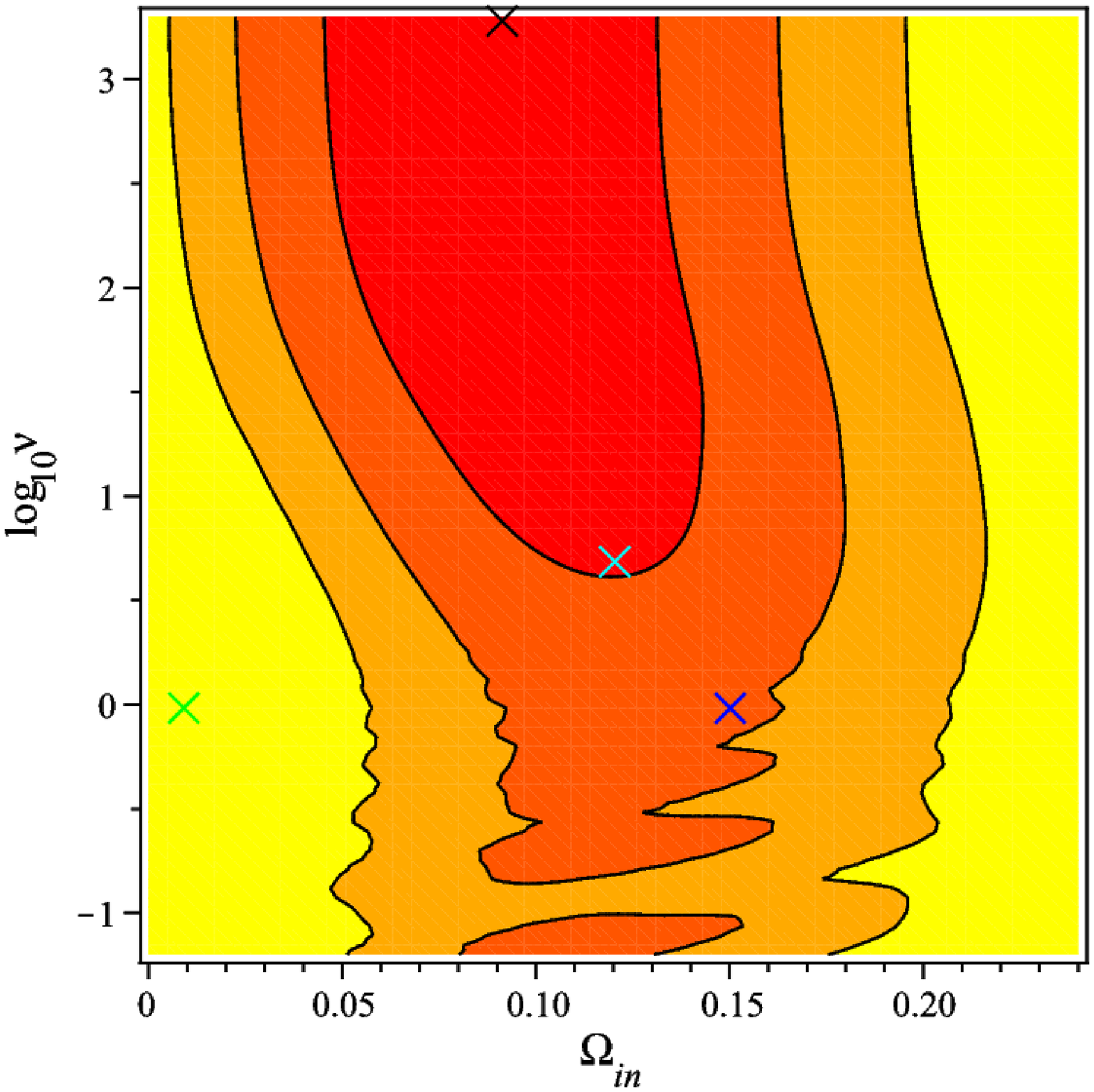,width=0.22\textwidth , clip} &
\epsfig{file=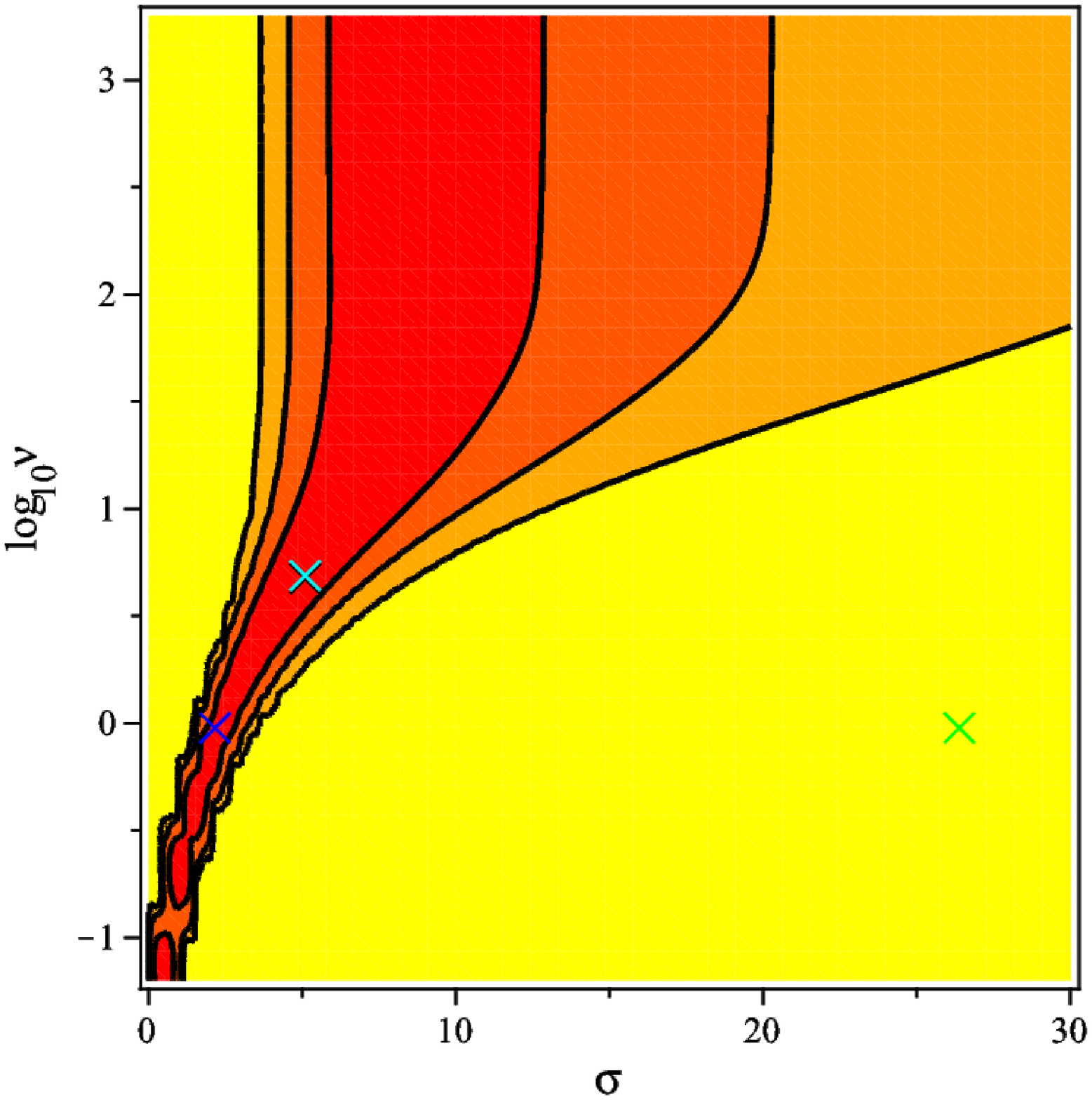,width=0.22\textwidth , clip}
\end{tabular}
\caption{Joint-parameter likelihood contours from SNIa+H constraints. Black crosses indicates the marginalized best-fitting values, blue crosses show the best-fitting to SNIa values, green crosses are the best-fitting to $H(z)$ values, and cyan crosses shows the best-fitting to SNIa+H values. See \S\ref{sec:H0} for a discussion of what fits to $H_0$ mean.}
\label{fig:likelihood_2D_CMBND}
\end{figure}
In Fig. \ref{fig:likelihood_2D_CMBND} we show the joint-parameter likelihood plots from SNIa+H constraints. The inner, red-filled regions represent the 68.3 per cent confidence level, the orange regions 95.4 per cent, and the off-yellow regions correspond to the 99.7 per cent level. The crosses indicate the positions of: the best-fitting to SNIa data (blue), the best-fitting to $H(z)$ data (green), the best-fitting to SNIa+H data (cyan), and the best-fitting marginalized (black), parameters.

Let us quickly discuss the degeneracies that exist between the parameters. The strongest degeneracies are between all the possible pairs of $\omegmin$, $\sigma$ and $\nu$, and can be interpreted as follows: in order to give the same $\chi^2$, if one wants a larger void (larger $\sigma$), then $\omegmin$ must decrease , and the void must be sharper at the origin (larger $\nu$), and vice versa. The degeneracies between those same parameters and $H_0$ are less obvious, but still present: emptier  and sharper  voids require a larger $H_0$, whereas a larger void  needs a lower $H_0$ in order to give similar $\chi^2$, and vice versa.
\begin{figure}
\centering
\begin{tabular}{cc}
\epsfig{file=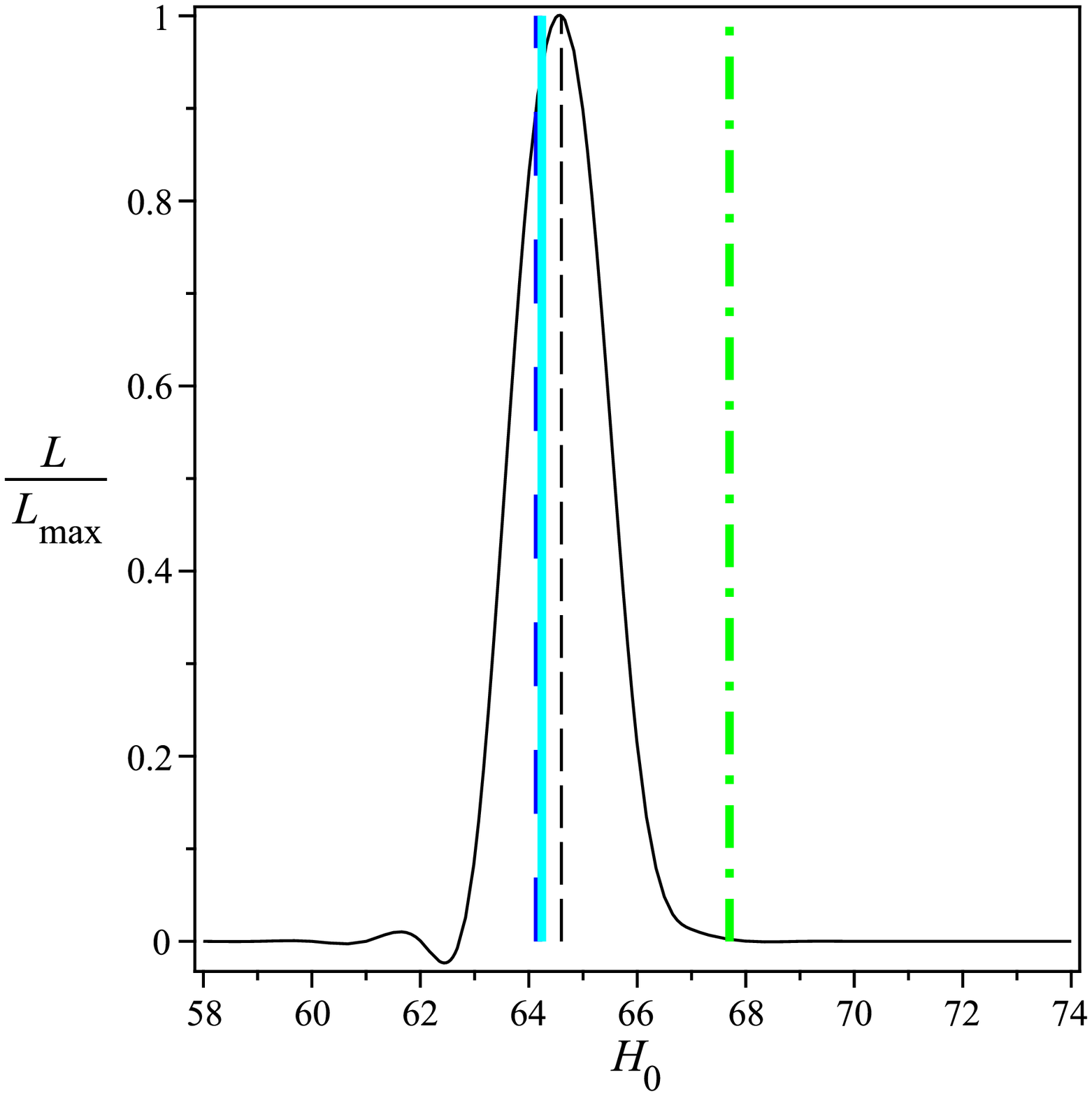,width=0.22\textwidth , clip} &
\epsfig{file=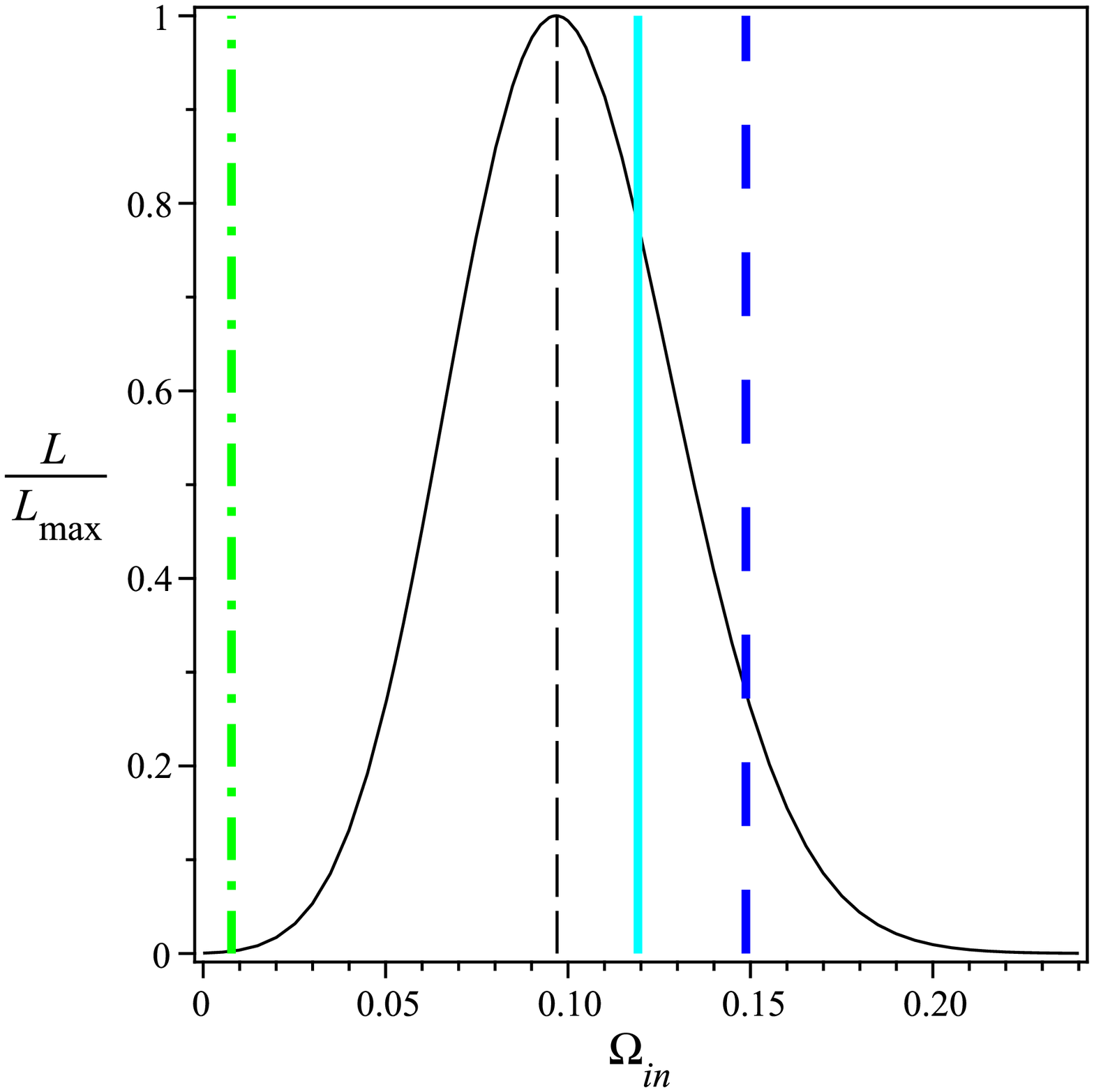,width=0.22\textwidth , clip}\\
\epsfig{file=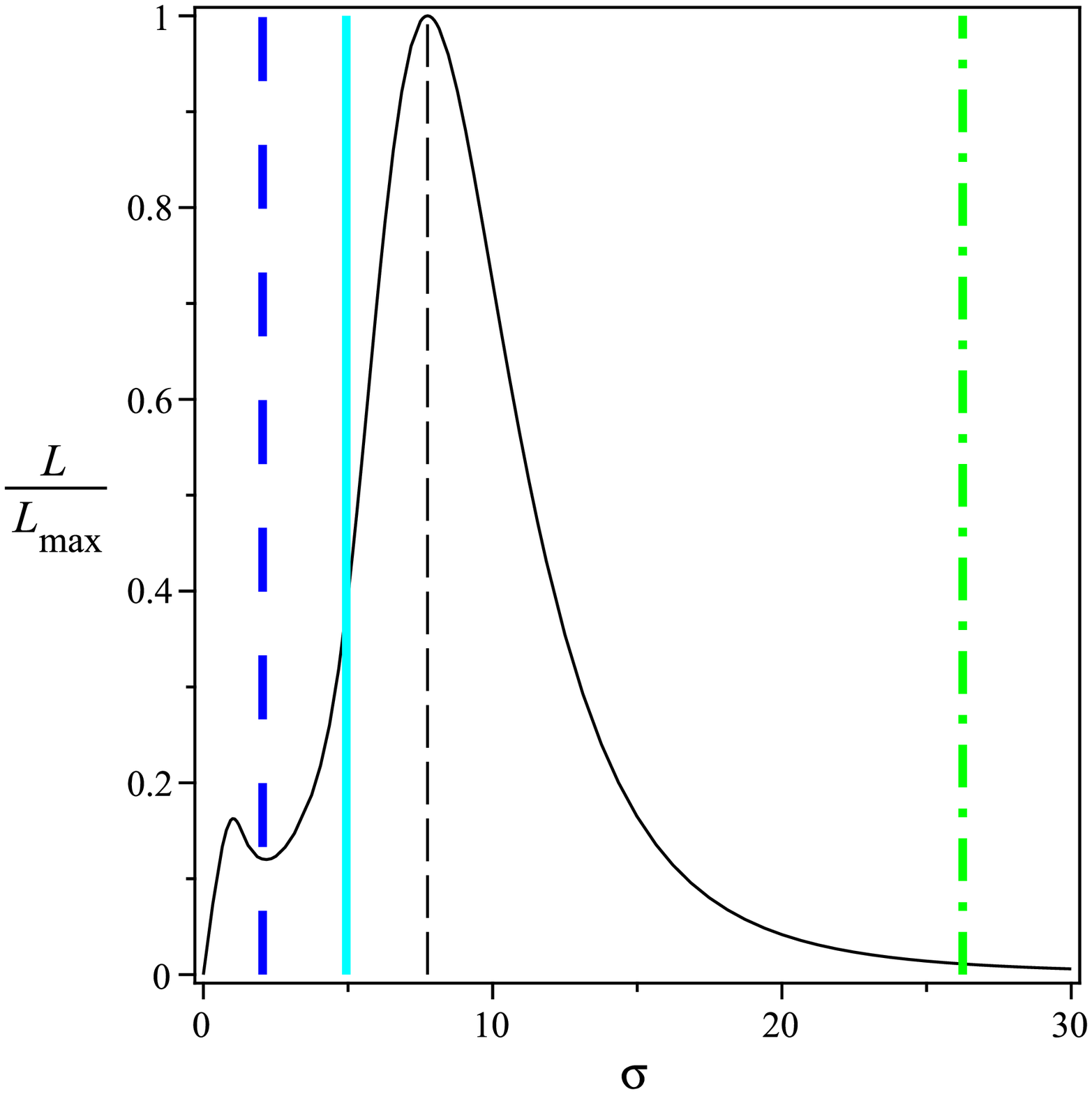,width=0.22\textwidth , clip} &
\epsfig{file=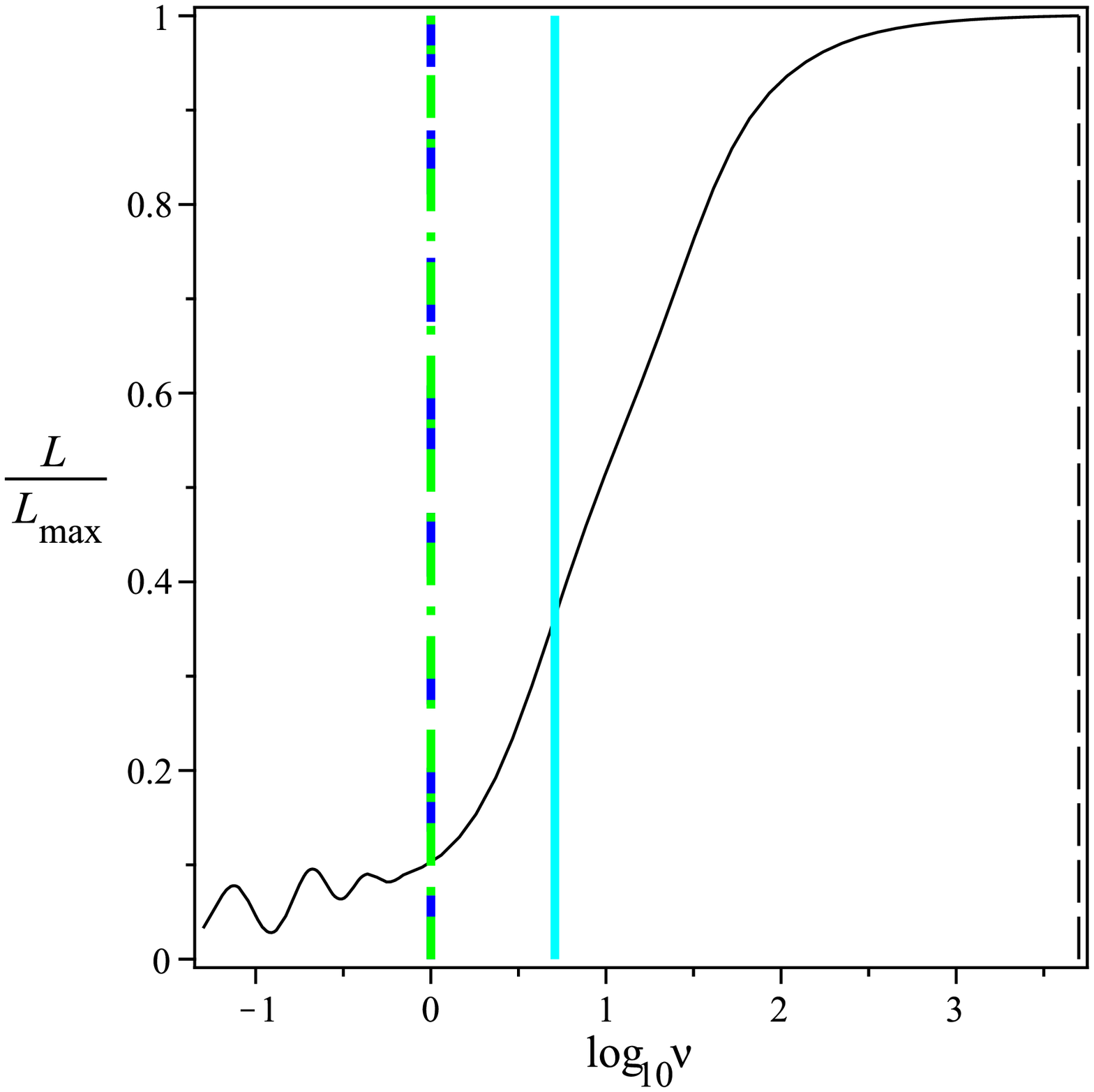,width=0.22\textwidth , clip}
\end{tabular}
\caption{1D likelihood distributions for each parameter from SNIa+H constraints. Black dashed lines corresponds to marginalized best-fitting values, while the blue, green and cyan lines correspond to the best-fitting to SNIa, $H(z)$ and SNIa+H values, respectively.}
\label{fig:likelihood_1D}
\end{figure}

In Fig.~\ref{fig:likelihood_1D} we show the one-parameter likelihood distributions for each parameter from SNIa+H constraints. The vertical lines correspond to: the best-fitting to SNIa (blue, space-dashed), best-fitting to $H(z)$ (green, dot-dashed), best-fitting to SNIa+H (cyan, solid) and marginalized best-fitting (black, dashed), parameters. Notice that $H_0$ and $\omegmin$ are fairly well constrained, whereas $\sigma$ and, even more so, $\nu$, are very poorly constrained individually owing to a degeneracy between them; taken together, however, the constraints in the $\sigma-\nu$ plane are reasonable. Note that $\nu$ is not constrained from above because $\nu=\infty$ is the $\LCDM$-void by construction.

\subsection{Which model is most preferred?}

According to the $\chi^2_\textrm{red}$ statistic, the void models considered here are, very slightly, favoured over $\LCDM$. However, there are more sophisticated techniques applicable to model selection, such as the (corrected) Akaike information criterion (AIC) \citep{akaike}
\ba
\textnormal{AIC} = \chi^2_{\textnormal{min}} + 2k + \frac{2k(k-1)}{N-k-1}\,,
\ea
where $k$ is the number of parameters and $N$ the number of data points, and the Bayesian information criterion (BIC) \citep{bayesian}
\ba
\textnormal{BIC} = \chi^2_{\textnormal{min}} + k\ln{N}\,.
\ea
A lower AIC or BIC implies a more favoured model. In Table \ref{table:infocrit} we summarize the AIC and BIC values for each void model as well as that of $\LCDM$.
\begin{table}
\caption{Information criterion for the best-fitting void models and that of $\LCDM$ model from SNIa+H constraints}
\centering
\medskip
\begin{tabular}{|c|cc|}
\hline
Model & AIC & BIC\\ [0.2ex]
\hline
1 & 329.70  &  345.68 \\[0.1ex]
2 & 327.78  &  339.77 \\[0.1ex]
3 & 327.60  &  339.60 \\[0.1ex]
4 & 327.84  &  339.84 \\[0.1ex]
5 & 327.89  &  339.89 \\[0.1ex]
$\LCDM$ & 330.52  &  338.53 \\[0.1ex]
\hline
\end{tabular}
\label{table:infocrit}
\end{table}
Using the AIC criteria we can see that all models are more-or-less equally favoured, whilst the BIC, which heavily penalises additional parameters, moderately disfavours void model \#1, in comparison to the remaining void models and $\LCDM$.

\subsection{Non-Concordance Diagnostics}
Finally, we consider quantities that can serve as useful conceptual tools for distinguishing FLRW/$\LCDM$ models from LTB models. Because the voids are really just toy models, constructing diagnostics from $\LCDM$ models allow us to visualize what our real constraints on $\LCDM$ are.

Consider the standard deceleration parameter in FLRW models which depends purely on the Hubble rate and the observable redshift:
\ba
q(z) = -1 + \frac{(1+z)}{H(z)}\frac{d}{dz}H(z)\ .
\ea
In testing for acceleration in FLRW models, we would like to ideally constrain this function directly from data~\citep{mortsellclarkson}. Now, using Eq.~(\ref{dtdz}), we can define an effective deceleration parameter in void models as
\ba\label{qeff}
q^{\textnormal{eff}}(z) = -1 + \frac{(1+z)}{\Hpar(z)}\frac{d}{dz}\Hpar(z)\ .
\ea
We anticipate, given the results of \citet{VFW}, that models containing a weak-singularity would have $q^{\textnormal{eff}}_0<0$, and models without the singularity would give $q^{\textnormal{eff}}_0>0$.
\begin{figure} 
\centering
\epsfig{file=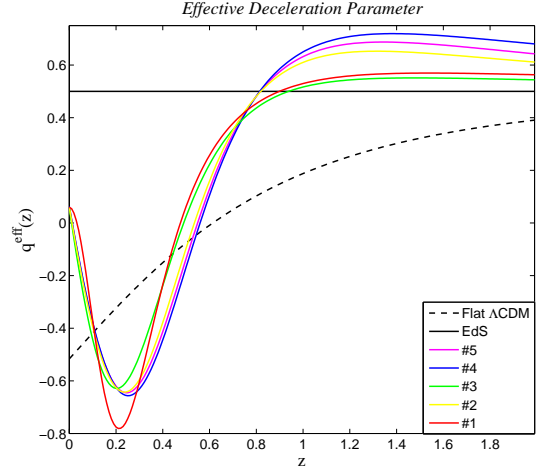,width=0.45\textwidth , clip}
\caption{Effective deceleration parameter for the best-fitting models to SNIa+H data for our 5 different void models, as well as that for $\LCDM$ and EdS. Note that for all these voids models we have an effective deceleration today.}
\label{fig:FIT2DATA_qz}
\end{figure}
In Fig.~\ref{fig:FIT2DATA_qz} we show the effective $q^\textnormal{eff}(z)$. These all have $q_0>0$ which is expected for smooth voids.
It is interesting to note that ~\citet{shafieloo} find $q_0>0$ for various classes of dark energy models within the FLRW paradigm when using the same datasets we consider here. In fact, it is worth comparing our $q^\textnormal{eff}(z)$ with the form they find when they fit only SNIa~-- they are qualitatively very similar.

\begin{figure}
\centering
\epsfig{file=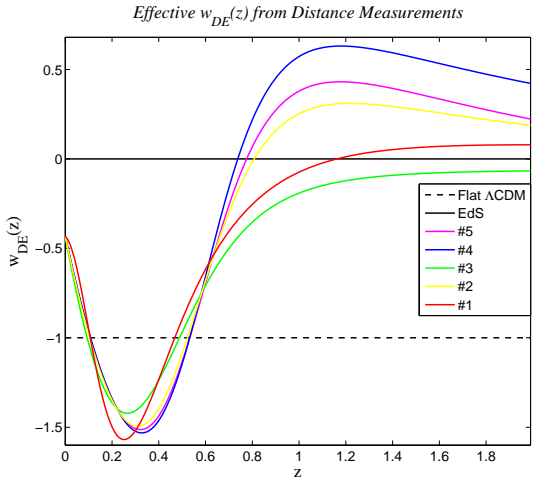,width=0.45\textwidth , clip}
\caption{Effective equation of state parameter from distance measurements for the best-fitting models to SNIa+H data for our 5 different void models, as well as that for $\LCDM$ and EdS.}
\label{fig:FIT2DATA_wz_DA}
\end{figure}

We may also consider an effective dark energy equation of state for the void models which we define as
\ba
\label{wdA}
w^{\textnormal{eff}}_{\textnormal{DE}}(z) =\frac{2(1+z) d_{\textnormal{c}}''  + 3d_c'\, }{3 \left[ H_0^2\omegam (1+z)^{3}{d_{\textnormal{c}}'}^{2} -1 \right ]d_{\textnormal{c}}'} \,,
\ea
where $d_{\textnormal{c}} = (1+z)\dA$ is the comoving angular diameter distance. This is the EOS which, in a flat FLRW dark energy model with energy density $\Omega_m$ gives the comoving angular diameter distance $d_{\textnormal{c}}(z)$. This gives us another nice way to visualize the differences between our voids and more standard dark energy functions used in the standard model. We show in Fig.~\ref{fig:FIT2DATA_wz_DA} this function for our best-fitting void models. Note the apparent phantom behaviour for moderate redshift.

Finally we consider two tests which have recently been proposed. The first is a test for the flat $\LCDM$ scenario:
\ba
\label{omrecon}
\omegam^{\textnormal{eff}}(z) =\frac{1-[H_0d_{\textnormal{c}}'(z)]^2}{[(1+z)^3-1][H_0d_{\textnormal{c}}'(z)]^2} =\frac{\left[ \Hpar(z)/H_0 \right ]^2 - 1}{(1+z)^3 - 1}\,,
\ea
should be constant if the concordance model is correct~\citep{litmus} (called $Om(z)$ in~\citet{sahni}), which implies that
\ba
\label{lz}
\mathscr{L} (z) &=& 2\left[(1+z)^3-1 \right ]H_0d_{\textnormal{c}}^{\prime \prime}(z) \nonumber \\ & & + 3(1+z)^2H_0d_{\textnormal{c}}^{\prime}(z)\left[1-(H_0d_{\textnormal{c}}^{\prime}(z))^2\right ]\,,
\ea
is zero in the concordance model~\citep{litmus}.

\begin{figure}
\centering
\epsfig{file=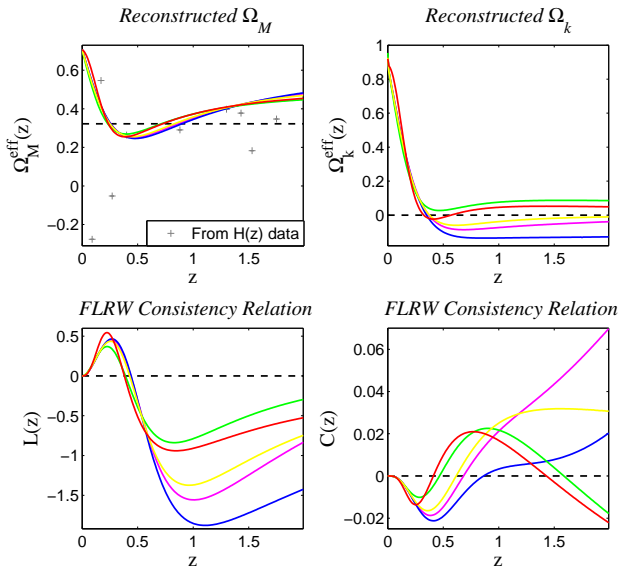,width=0.52\textwidth}
\caption{Top: Effective $\omegam^\text{eff}(z)$ and $\omegak^\text{eff}(z)$ for our 5 best-fitting void models to SNIa+H data, along with that of the $\LCDM$ model. For the case of $\omegam^\text{eff}(z)$, we have placed the $H(z)$ data used in fitting the models, without the error bars (which are rather large).  Bottom: the corresponding second-derivative tests $\mathscr{L}(z)$ and $\mathscr{C}(z)$.}
\label{fig:conctests}
\end{figure}

The second test is a much more general test of a FLRW geometry~\citep{CBL}.
\ba
\label{okrecon}
\omegak^{\textnormal{eff}}(z) = \frac{\left[d_{\textnormal{c}}(z)\Hpar(z)\right ]^2-1}{\left[d_{\textnormal{c}}(z)H_0\right ]^2}\,,
\ea
should be constant for any FLRW dark energy model, and
\ba
\label{cz}
\mathscr{C} (z) &=& 1 + \Hpar(z)^2\left[d_{\textnormal{c}}(z)d_{\textnormal{c}}^{\prime \prime}(z) - {d_{\textnormal{c}}^{\prime}(z)}^2\right ] \nonumber \\ & & + \Hpar(z)\Hpar^{\prime}(z)d_{\textnormal{c}}(z)d_{\textnormal{c}}^{\prime}(z)\,,
\ea
should be zero. The utility of these tests is that they can be used independently of any model at all. Considering these functions in our void models helps us visualize the difference from FLRW in terms of observable functions. This is shown in Fig.~\ref{fig:conctests}. Note that for $\omegam^\text{eff}(z)$, our best-fitting voids produce a curve which is very similar to that found by~\citet{shafieloo} considering FLRW dark energy models.

\section{Discussion}
\label{sec:conc}

In this work, we have discussed LTB void models as solutions to the dark energy problem, and have considered how to constrain them with cosmological data.
Of particular interest is the steepness of the void near the centre of symmetry, as well as the need to combine independent data sets in order to distinguish voids from $\LCDM$.

As far as SNIa and distance data are concerned, it is often claimed that to fit the data a spiky void near the centre is required; this is not the case. To reproduce the $\LCDM$ distance modulus \emph{exactly}, one needs a void with a point at the origin; if one fits the actual data, and not a prediction of the $\LCDM$ model, smooth voids are perfectly compatible with observations, and their likelihoods make them indistinguishable from a $\LCDM$ scenario. By introducing a new parameterization we can smoothly parameterize away from the ``$\LCDM$-voids'', to test specifically how steep the void would have to be.

Moreover, we have included $H(z)$ data in the hope of removing the degeneracy between a class of voids and the $\LCDM$ models. Unfortunately, current uncertainties on the age measurements do not yet allow for such a discrimination between the models, with only the void model with an additional parameter being moderately disfavoured by the BIC compared to $\LCDM$. All other models are equally favoured by current datasets.

On the other hand, a careful look at the joint constraints from SNIa and $H(z)$ data provides evidence for a robust estimation of the void size through the full width at half maximum. Indeed, in agreement with previous results, we have found that the size of the void is of order 6 Gpc. This estimation appears to be fairly independent on the parameterization of the void profile.

Finally, we have illustrated how our void models look when viewed in terms of standard FLRW functions. For example, the voids, interpreted through the lens of a flat dark energy fluid give a $w(z)$ which has phantom behaviour at intermediate redshifts, and a deceleration parameter which is positive for low redshift. These profiles may be usefully considered as signatures of voids. Furthermore, we have also shown profiles for $\Omega_m^\text{eff}(z)$, which is constant only for flat $\LCDM$, and  $\Omega_k^\text{eff}(z)$ which is constant in any FLRW model whatsoever. These will be very useful diagnostics in future studies of dark energy because they allow us to measure deviations from flat $\LCDM$ and FLRW models respectively in a model independent way. By seeing what they look like in void models, we understand further the range of non-concordance behaviour possible in these diagnostics. It is rather interesting to note, in fact,  that the effective deceleration parameter, as well as the effective matter density predicted by LTB models look very similar to previous estimations of these quantites relying on a parameterization of the equation of state of Dark Energy \citep{shafieloo}.

As a final remark, recent constraints on non-standard cosmologies were presented in \citet{sollerman} using the first-year Sloan Digital Sky Survey-II (SDSS-II) supernova results \citep{sdssII} in combination with other data, such as the latest BAO and CMB measurements. In particular, using a simple Gaussian void profile allowing $\omegmout$ to vary, \citet{sollerman} demonstrate that with this new dataset LTB models are still capable of fitting the SNIa-only data better than the $\LCDM$ model.

However, \citet{sollerman} do not find evidence that the additional d.o.f. they allow when compared to our Gaussian profile is supported by the latest dataset. This dataset contains new SNIa, which now fill in the gap around $z \sim 0.2$, but lack the recent low redshift supernovae that are included in the Constitution dataset. We note that their conclusion that LTB models are a worse fit than $\LCDM$ are as a result of using BAO/CMB data which we do not use here for the reasons discussed above. Thus, we expect that their new SNIa data will affect our conclusions only marginally.

This work strongly suggests the importance of testing LTB models with further observations such as CMB and BAO, jointly with SNIa and age measurements, in order to discriminate between models. Such analyses, however, will rely on a careful implementation of structure formation and CMB emission in an LTB background as initiated in~\citet{pertpaper}. This will be the topic of forthcoming studies.

\section*{Acknowledgments}

We thank Timothy Clifton for very helpful discussions. CC is supported by the National Research Foundation (South Africa). MS is supported by SKA / KAT (South Africa). JL is supported by the Claude Leon Foundation (South Africa). SF is funded by the National Astrophysics and Space Science Programme (South Africa) and the SKA. The authors would also like to thank Seshadri Nadathur for pointing out minor corrections with a previous version of this manuscript.

\label{lastpage}

\end{document}